\documentclass[a4paper,11pt]{article}

\usepackage{jcappub}
\usepackage[T1]{fontenc}
\usepackage{multirow}
\usepackage{jabbrv}
\usepackage{bm}

\title{Constraining the mass-concentration relation through weak lensing peak function}

\author[a]{R. Mainini,}
\author[b]{A. Romano}

\affiliation[a]{Physics Department, University of Milano--Bicocca, \\
Piazza della Scienza 3, I20126, Milano, Italy }
\affiliation[b]{INAF- Astronomical Observatory of Rome, \\
           via Frascati 33, I00044, Monte Porzio Catone (RM), Italy}
\emailAdd{roberto.mainini@mib.infn.it}

\begin{abstract}
{Halo masses and concentrations have been studied extensively, by means of 
N--body simulations as well as observationally, during the last decade. 
Nevertheless, the exact form of the mass--concentration relation is still widely debated.
One of the most promising method to estimate masses and concentrations 
relies on gravitational lensing from massive halos. 
Here we investigate the impact of the mass--concentration relation on halo peak
abundance in weak lensing shear maps relying on the aperture mass
method for peak detections. 
After providing a prescription to take into account the concentration
dispersion (always neglected in previous works) in peak number counts
predictions, we assess their power to constrain the
mass--concentration relation by means of Fisher matrix technique.
We find that, when combined with different cosmological probes, peak
statistics information from near--future weak lensing surveys provides an
interesting and complementary alternative method to lessen the long standing 
controversy about the  mass-concentration relation. }
\end{abstract}

\begin{document}

\maketitle
\flushbottom

\section{Introduction}

Weak gravitational lensing (WL) has emerged over the last decade as one of
the most promising methods for testing cosmology and gravity, and 
unveiling the nature of Dark Energy (DE) and Dark Matter (DM).
It relies on the accurate measurement of the
small shape distortions of background galaxies due to the bending of light 
by intervening matter distribution. 
The smallness of these distortions requires high-quality data  and a statistical approach
so that the lensing shear signal needs to be measured on 
a large number of sources. 

A number of surveys (e.g. GaBoDS \cite{Hett07}, 
CFHTLS\footnote{http://www.cfht.hawaii.edu/Science/CFHLS}) have already shown 
the WL capacity for constraining 
cosmological models through cosmic shear measurements (see e.g. 
\cite{Hoekstra06,Semboloni06,Hett07}).  
Several other, more ambitious, surveys 
both ground-- 
(KIDS\footnote{http://kids.strw.leidenuniv.nl}, 
PanSTARRS\footnote{http://pan-starrs.ifa.hawaii.edu}, 
DES\footnote{http://www.darkenergysurvey.org}, 
LSST\footnote{http://www.lsst.org})
 and space-based 
(Euclid\footnote{http://www.euclid-ec.org}, 
WFIRST\footnote{http://wfirst.gsfc.nasa.gov}) are being performed or planned. 
They will map hundreds of millions of galaxy redshifts and billions
of galaxy images. For instance, 
the Euclid mission \cite{euclid1,euclid2} aims to map half of the sky in imaging as well as
in broad-band spectroscopy up to a redshift 
$z \sim 2-3$ with a median redshift 
of the order of unity.
The information contained in these data 
will permit to measure the matter clustering with 
unprecedented accuracy.

Among the WL probes, it has been shown that
the abundance of peaks in shear maps 
is a sensitive probe of cosmology \cite{kruse99}.  
This was tested by \cite{Reblinsky99} and \cite{Dietrich10}
using ray--tracing in N--body simulations. 
Shear peaks are regions with high signal--to--noise ratios, $snr$, associated 
to massive halos (or galaxy clusters) or produced by the alignment
of smaller mass concentrations along the line--of--sight. 
Their number and spatial distribution, therefore, carry information about 
the underlying cosmology through fundamental parameters as the
total matter density of the universe $\Omega_m$, the normalization of 
the power spectrum $\sigma_8$, and the evolution of the DE equation of state $w(z)$. 

Unlike other available techniques for detecting  massive halos 
(e.g. optical and X--ray identification, Sunyaev-Zeldovich effect, etc), 
WL does not require any assumption on their dynamical or evolutionary state.
Its advantage is to be only sensitive to the mass along the line--of--sight 
offering, in principle, the opportunity to construct mass--selected halo samples,
which can be compared directly with theoretical predictions (e.g. N--body
simulations) without assuming any mass--observable relation.   
Nevertheless, that WL provides truly mass--selected halo samples 
it is not strictly true.
Large scale structures (LSS), into which 
massive halos are embedded, contribute 
to the lensing signal. Usually, these  projection effects add noise to 
the halo signal or can result in false positive peaks in the shear map
\cite{Hoekstra01,Hoekstra03,Hoekstra11}.
Further, intrinsic ellipticities of galaxies introduce an irremovable noise.
All these sources of noise then compromise purity and completeness of 
halo samples selected by WL.

Various theoretical aspects of WL halo detection have been widely investigated 
by a number of authors 
\cite{hamana04,Dietrich10,marian06,maturi10,maturi11,cardone}.
The search for shear peaks in data is exemplified in some works 
\cite{hett05,dahle06,wittman06,gavazzi07,schirmer07,miyazaki07,berge08,abate09,shan11}
although, on the observational side, halo lensing studies have been mainly 
focused on mass determination and halo properties, e.g. density profile and concentration \cite{okabe10,israel10}.

A basic method for shear peak detection, to which we will refer in the 
following, is provided by the aperture mass ($M_{ap}$) technique proposed by 
\cite{S96}. $M_{ap}$ is a measure of the lensing signal in a shear 
map smoothed with a suitable filter.
The method then relies on the usual approach of searching for points of 
local maximum (peaks), with $snr$ higher than a given threshold, 
in the smoothed map. 

Here, we investigate the impact of the halo concentration on $M_{ap}$ 
and shear peak counts, as compared to that of some cosmological parameters 
(i.e. $\Omega_m$ and $\sigma_8$).
As already noted by \cite{king11} and detailed in the following,
the shear peak abundance, 
is quite sensitive to halo concentration and its relation with the halo mass.
This seems to point out  peak counts as a promising and 
complementary method to delineate the, still debated, relation between 
halo mass and concentration ($M-c$ relation). Our main interest here
is to further deepen this point quantifying the power of peak function
to constrain the $M-c$ relation. 
 
Halo concentrations have been studied extensively, by means of N--body 
simulations as well as observationally, during the last decade.
The general trend is that the mean halo concentration shows a strong 
correlation with the halo mass. Typically, it declines with increasing the 
mass and redshift in accordance with the idea that the central density of 
halos reflects the mean density of the universe at the time of their formation. 
Therefore, halos collapsing earlier are expected to be denser
than the more massive halos collapsing later. 

However, there is enough uncertainty on the exact form 
of the $M-c$ relation.
Beside discrepancies between theoretical predictions 
and observations, differences are found even when comparing simulation 
results of different groups as well as different sets of observations 
\cite{B01,eke01,Comerford07,neto07,gao08,duffy08,maccio08,mandelbaum08,oguri09,okabe10,klypin11,prada11,meneghetti13}.

One of the main predictions of N--body simulations is that the
density profile of halos assembled hierarchically and close to 
virial equilibrium, can be well approximated
by a Navarro, Frenk \& White (NFW) profile \cite{NFW}, 
regardless of halo mass and the details of the cosmological model, 
all the cosmological informations being contained in correlation between the 
parameters of the NFW profile, i.e. in the $M-c$ relation. 
The median $M-c$ relation predicted by simulations is well 
described by a power--law \cite{dolag04,gao08,duffy08,maccio08,zhao09}.

However, the variety of individual halo aggregation histories
causes a scatter in concentration which
can be modeled by a log--normal distribution with a variance ranging from 
$\sim 0.15$ up to $\sim 0.30$ 
depending on the degree of relaxation of the halo \cite{Jing00}.
On the other hand, the redshift evolution of the concentration is 
less established. Recent results seem to favor a weaker redshift 
dependence than previously supposed \cite{duffy08,maccio08,MC11}.

Probably, the best current results from simulations are 
those by \cite{klypin11} and \cite{prada11}. 
They found larger concentrations than those reported in \cite{maccio08} and \cite{duffy08}, with a
reasonable agreement for galaxy--size halos ($\sim 10-15$\% difference) but 
substantially larger values ($\sim 40-50$\%) for cluster--size halos, more 
compatible with recent X--ray and kinematic observations. Furthermore, these 
high--resolution large-volume simulations ({\it Bolshoi} and {\it MultiDark}, see http://www.mutidark.org)
seem to indicate a $M-c$ relation more complex than previously conceived showing a novel feature: 
at high redshift, concentration first declines with increasing mass, 
then flattens and  increases slightly at higher masses. 

On the observational side, many different methods have been used to study 
masses and concentrations resulting in an even more controversial picture
and making the comparison between observations and 
theory somewhat ambiguous.      
 
Gravitational lensing is one of the most promising 
methods to estimate mass and concentrations.
Strong and weak lensing have been used individually or in combination.
More classical methods relies on kinematic tracers, e.g. galaxies 
(\cite{rines06,WL10} and references therein)
and the hydrostatic analysis of X--ray intensity profile of halos.
However, all these methods are susceptible to bias effects and are 
presumably affected by systematics not fully understood 
and/or modeled, yielding discrepant results.

For example, \cite{mandelbaum08} used the stacked weak lensing signal 
from galaxies, groups and clusters in the Sloan Digital Sky Survey (SDSS)  
finding a $M-c$ relation slope consistent with the simulations 
but a $2\sigma$ lower normalization. 
\cite{oguri09} performed a combined weak and strong lensing 
analysis of a sample of clusters,
reporting concentrations with a $7\sigma$ excess 
above the simulation predictions.
\cite{okabe10} performed a weak lensing analysis
of X--ray selected clusters
reporting a slope somewhat steeper than in simulations 
although with an error of $\sim 50$\%. 
Other studies based on lensing and X--ray data can be found in
\cite{Comerford07,buote07,SA07,johnston07,broadhurst08,vikhlinin09,ettori11}.

Despite the qualitative agreement with simulation predictions, the general 
picture which emerges from observations can be summarized briefly as follows.
In almost all cases, the observed slope of the $M-c$ relation is
consistent with or steeper than theoretical predictions. 
Moreover, strong lensing measures of massive clusters and X--ray analysis 
give a normalization factor higher than predicted by simulations, 
strong lensing concentrations being systematically 
larger than X--ray concentrations. 
On the other hand, weak lensing methods seem to point out a
normalization lower than that found in simulations.

For what concerns the lensing measurements, the origin of these 
discrepancies should be searched in some orientation and shape biases 
\cite{oguri05,sereno11,giocoli}.
It has been shown that neglecting halo triaxiality can lead to over-- and under--estimates of 
a factor of $2$ in concentrations and up to $50$\% in halo mass other than underestimation
of statistical uncertainties \cite{corless09}. In addition, as already
outlined above, projection of structures along the line--of--sight can result
in apparently high concentration (e.g. \cite{KC07}). 

Further complications are due to baryonic processes.
Although baryon physics is not expected to drive 
the structure formation process on very large scales, 
its effects are likely to be important for low--mass halos 
or in the central region of larger objects. 
It has been shown that baryonic feedback and cooling
can alter halo profiles. In particular, baryonic cooling
(see \cite{gnedin11} and reference therein)
can be responsible for the excess in concentration observed
in groups and low-mass clusters \cite{sereno10,fedeli11}.

Dynamical and X--ray techniques assume the halos are in virial and/or
hydrostatic equilibrium. 
Most of real halos are not relaxed and have 
complex substructures making measurements difficult to interpret. 
Mergers, active galactic nuclei, cosmic rays, 
magnetic fields, turbulence and bulk motion of the gas can compromise 
the hydrostatic equilibrium leading to underestimates of X--ray masses
\cite{evrard96,dolag05,rasia06,nagai07,rasia13}.

Although recent hydrodynamical simulations include a number of these
non--gravitational processes, they 
are very difficult to model. Furthermore, many of them take place
at scales too small to be resolved by simulations 
demanding radical simplifications and approximations.
 
It is also worth noticing that the above results were derived in the contest 
of the standard cosmological model ($\Lambda$CDM). Discrepancies can also
arise if the data--model comparison has been made for the wrong cosmology.
A number of studies have shown that alternative DE cosmologies or modified gravity models can lead to
a different evolution of matter perturbations affecting the halo 
concentrations. For example, scalar field DE models with an   
earlier structure formation resulting in more concentrated halos have been 
considered in \cite{dolag04,io1,io2,baldi,baldi2,boni}.

The paper is organized as follows: the $M_{ap}$ method for shear peak detection
is reviewed in Section \ref{Map} while in Section \ref{PF} 
we investigate the impact of the $M-c$ relation on the probability 
distribution function of $M_{ap}$ and provide a prescription to analytically
calculate it taking into account the dispersion in $M-c$ relation.
The impact on peak number counts is then investigated and compared to that 
of $\Omega_m$ and $\sigma_8$. In Section \ref{fisher} we perform a Fisher matrix 
analysis in order to assess the capability of upcoming weak lensing survey
in constraining the $M-c$ relation by means of peak counts.
Section \ref{concl} is devoted to discussion and conclusions.

\section{Aperture Mass}\label{Map}

The gravitational field of a sufficiently high matter concentration 
causes a coherent shape distortion of the images of faint background galaxies. 
Measurements of these shear distortions, would thus permit to 
detect massive halos by searching for peaks in wide--field shear maps. 
To this aim, different WL methods for halo detection, based on
linear filtering techniques,
have been developed in the last decade
\cite{S96,hamana04,schirmer,maturi05,HS05}  (for alternative 
approaches see \cite{KS93,KSB95,JVW00}).

Here, we use the aperture mass, $M_{ap}$, \cite{S96}
as peak finder.
$M_{ap}$, is the projected mass 
distribution, or convergence $\kappa$, smoothed
on the angular scale $\vartheta_0$:
\begin{equation}
M_{ap}(\vartheta_0) = \int d^2 \theta \kappa({\bm \theta}) U(|{\bm \theta}|) 
\label{apmass}
\end{equation}
where ${\bm \theta} = (\vartheta \cos \phi,\vartheta \sin \phi)$ 
and $U(\vartheta)$ is a compensated filter function, 
i.e. $\int_0^{\vartheta_0} d \vartheta \vartheta  U(\vartheta)=0$, which vanishes
for $\vartheta > \vartheta_0$. The main 
advantages in using compensated filters are that $M_{ap}$ is not 
influenced by mass--sheet degeneracy and 
the possibility of expressing
 $M_{ap}$ in terms of the (observable) tangential shear $\gamma_t$:
\begin{equation}
M_{ap}(\vartheta_0) = \int d^2\theta \gamma_t({\bm \theta}) 
Q(|{\bm \theta}|)
\label{mapg} 
\end{equation}
where $\gamma_t({\bm \theta})=-Re[\gamma(\bm \theta)e^{-2i\phi}]$ is the tangential component of the shear at position ${\bm \theta}$
and $Q$ is a filter function related to $U$ by:
\begin{equation}
 Q(\vartheta) = \frac{2}{\vartheta^2} \int_0^\vartheta d\vartheta' \vartheta' U(\vartheta') - U(\vartheta)  
\end{equation}

\subsection{Filter function}

Several filter functions $Q(\vartheta)$ have been considered depending on their
specific application in WL studies (cosmic shear, halo searching, etc.).
The ideal choice for radially symmetric halos, would be a filter 
profile proportional to the shear profile, i.e. $Q(\vartheta) \propto 
\gamma(\vartheta)$ thus to maximize the signal--to--noise 
ratio $snr$ associated to halo detections as shown in \cite{S96}.
Some examples of filters can be found in 
\cite{SWJK} (polynomial filter), \cite{hamana04} (Gaussian filter),
\cite{maturi05} (optimized for NFW halo and LSS noise suppression), 
\cite{gruen} (non--parametric filter). 

To our aim, a reasonable choice is the filter introduced 
by \cite{schirmer} and optimized for NFW halos, approximating 
their shear signal with a hyperbolic tangent:
\begin{equation}
Q(x)=\left[\pi\vartheta_0\left(1+e^{a-bx}+e^{-c+dx}\right)\right]^{-1} 
\frac{{\tanh} \left({x}/{x_c}\right)}
{({x}/{x_c})}
\end{equation}
where $x=\vartheta/\vartheta_0$. 
The values $a=6$, $b=150$, $c=47$ and $d=50$ are chosen in order to have 
exponential cut--offs at small and large radii while $x_c=0.1$ is a good 
choice for the filter profile slope as empirically shown by \cite{hett05}. 
The related filter $U(\vartheta)$ is obtained following the procedure 
described in \cite{maturi10}. In the following we will always set 
$\vartheta_0=7'$.
We have checked that this choice is the best compromise which maximizes
the number of detections over the redshift range of our interest.

\subsection{Halo profile and  mass--concentration relation}

In order to compute the convergence $\kappa$
of cluster-sized halos  entering in (\ref{apmass}),
we model their density profile by a Navarro, Frenk \& White (NFW) profile:  
\begin{equation}
\rho(r)=\frac{\rho_s}{(r/r_s)(1+r/r_s)^2}
\end{equation}
The finite extent of halos is taken into account by truncating the 
profile at the radius $r_{200}$, within which the average density, 
$\Delta_{200} ~\rho_{cr}$, is $200$ 
times the critical density of the universe $\rho_{cr}=3H^2/8\pi G$
($H$ and $G$ being the Hubble parameter and the Newton's gravitational
constant respectively).

The scale radius $r_s$ is related to $r_{200}$ through the concentration 
parameter $c_{200}=r_{200}/r_s$ giving:
\begin{equation}
\frac{\rho_s}{\rho_{cr}}=\frac{1}{3}\Delta_{200}f_{200}
\end{equation}
where:
\begin{equation}
f_{200}=c_{200}^3 \left[\ln(1+c_{200})-\frac{c_{200}}{1+c_{200}}\right]^{-1}
\end{equation}
The halo mass is then defined as:
\begin{eqnarray}
M_{200} & = &\frac{4\pi}{3}\Delta_{200}\rho_{cr}r_{200}^3
  =  4\pi\rho_sr_{200}^3
f^{-1}_{200}
\end{eqnarray}

A number of heuristic models for the median mass--concentration relation, 
calibrated against simulations, have been suggested 
\cite{NFW,B01,eke01,zhao09}.
Here we assume the common simple relation: 
\begin{equation}
c_{200}(M_{200},z) = c_0 \left[\frac{M_{200}}{M_0} \right]^\alpha
g(z)
\label{c200}
\end{equation}
with the concentration growth factor, $g(z)$, as proposed by \cite{maccio08}
and we set $M_0=10^{14}h^{-1} M_\odot$.
Their model is a modification of the commonly used Bullock model \cite{B01}, 
based on the simple assumption that the characteristic 
density of the halo, $\rho_s$, remain constant after the halo forms. 
This results in a concentration growth factor: 
\begin{equation}
g(z)= \left[\frac{H_0}{H(z)}\right]^{2/3}=
 \left[\frac{\rho_{cr,0}}{\rho_{cr}(z)}\right]^{1/3}
\label{cgrowth}
\end{equation}
rather than $g(z)= 1/(1+z)$ of the original Bullock model. 
The normalization of the $M-c$ relation
$c_0$ is thus defined as the
concentration $c_{200}$ of a halo with mass $M_{200} = 10^{14}h^{-1} M_\odot$ 
at $z=0$.
Note that (\ref{cgrowth})
seems more consistent with the 
idea that halo concentrations are related to the ratio of the present
background density, $\rho_{cr,0}$, and that at the halo formation epoch. 
It reduces to the Bullock model
in the case of Einstein--De Sitter universe and is able to reproduce the 
$M-c$ relation,
in a $\Lambda CDM$ universe, over a wide range of masses 
($10^{10}h^{-1}M_{\odot}\lesssim M \lesssim 10^{15}h^{-1}M_{\odot}$) 
probed by simulations. Its scaling with $\rho_{cr}$, however, would suggest
that it could be more suitable than the Bullock model 
in describing the mass--concentration evolution in cosmologies 
different from $\Lambda CDM$. For simplicity, we assume a constant
slope $\alpha$ over the mass range considered although simulations
indicate a slight deviation at higher masses.

The convergence for a truncated NFW profile then reads \cite{hamana04}:

\begin{equation}
\kappa(\vartheta,z_h, z_g)=\frac {8\pi G}{c^2}\rho_s r_s
\frac{D_{h}D_{hg}}{D_{g}}f(x),
\end{equation}

where $x=\vartheta/\vartheta_{s}$, $\vartheta_{s}=r_s/D_h$, $D$
denotes the angular diameter distance, the subscripts $_h$ and $_g$
stand for lens and source galaxies respectively. Finally, the function
$f$ is given by:

\begin{displaymath}
f(x)=\left\{
\begin{array}{lr}
-\frac{\sqrt{c_{200}-x^2}}{(1-x^2)(1+c_{200})}+(1-x^2)^{-3/2}\,{\rm arccosh} \frac{x^2+c_{200}}{x(1+c_{200})} & x<1\\ \\
\sqrt{c^2_{200}-1} ~\frac{(2+c_{200})}{3(1+c_{200})^2} & x=1\\ \\
-\frac{\sqrt{c_{200}-x^2}}{(1-x^2)(1+c_{200})}-(x^2-1)^{-3/2}\,\arccos \frac{x^2+c_{200}}{x(1+c_{200})} & 1<x\leq c_{200}\\ \\
0 & x>c_{200}
\end{array}
\right.
\end{displaymath}

\subsection{Galaxy survey}

If the source galaxies are distribute in redshift, the convergence $\kappa$
entering in (\ref{apmass}) is the convergence averaged over the normalized 
source--redshift distribution $p(z_g)$ ($z_g$ being the redshift of the 
source galaxies). For a NFW halo at a redshift $z_h$
with convergence $\kappa(\vartheta,z_h,z_g)$ then we have:
\begin{equation}
\kappa(\vartheta)=\int_{z_h}^{\infty}dz_g p(z_g) \kappa(\vartheta,z_h,z_g)
\end{equation}
Here, we assume a distribution $p(z)$ of the standard form:
\begin{equation}
p(Z)=\frac{\beta}{z_0 ~\Gamma\left(\frac{a+1}{b}\right)} 
Z^a e^{-Z^b}
\label{gal}
\end{equation}
where $Z=z/z_0$ and we set $a=2$, $b=1.5$ and $z_0=0.6$. 
We further assume an ellipticity dispersion of the sources 
$\sigma_\epsilon=0.3$, a galaxy number density $n=30 ~gal/arcmin^2$ and 
a survey area of 15000 $deg^2$.
With these values (\ref{gal}) provides a galaxy distribution as expected 
from the Euclid survey, with a mean redshift 
$z_{mean} = 0.9$ \cite{marian11,euclid1}.

\section{Peak function}\label{PF}

In real observations, the measured value of $M_{ap}$ will differ from the 
real one due to several effects. Statistical noise and halo shape induce 
scatter and bias in $M_{ap}$ and hence in the resulting shear peak counts.
A detailed analysis on how noise and the halo structure affect the 
probability distribution function (PDF) of $M_{ap}$ 
and the peak counts was performed in \cite{hamana04,hamana12}
(and references therein, see also \cite{fan} and \cite{bahe})
employing both analytical description of halos and mock data of 
WL survey generated from numerical simulations. 
Statistical noise due to intrinsic ellipticities, survey shot noise and 
projection effects of large scale structure (LSS) along the line--of--sight,  
affect the peak heights and generate spurious peaks in the shear maps resulting 
in an excess in real counts over the theoretical predictions. 

On the other hand, the dark matter distribution in real halo is not 
spherically symmetric but highly elongated \cite{jingsuto} so that 
deviations from the universal NFW density profile are expected.
Furthermore, simulation results tell us that halo concentrations are 
log--normally distributed among halos with a given mass, 
with variance $\sim 0.15-0.3$ \cite{Jing00}.

All these effects have a large impact on WL measurements of halos and induce 
scatter and bias in the peaks heights therefore affecting the peak counts.
For simplicity, we do not consider here deviation from sphericity and assume
halos to have NFW profiles while we only focus on the effects due to 
statistical noise and, in particular, concentration scatter.  
Impact of halo triaxiality on WL peak searches was accurately studied in 
\cite{hamana12}.

In order to predict the expected abundance of halo peaks we need to know 
the PDF of $M_{ap}$ and the halo mass function. As a fiducial model we assume
a $\Lambda$CDM cosmology with cosmological parameters set to the best fit 
values estimated from nine-years WMAP data in conjunction with the Supernova 
Legacy Survey three--year sample (SNLS3) and baryon acoustic oscillations (BAO)
measurements \cite{wmap9}. Fiducial cosmological parameters are
summarized in table \ref{par} ($\Omega_b$, $H_0$, $\tau$ and $n_s$ denoting the
baryon density parameter, the present value of the Hubble parameter, the 
optical depth to reionization and the spectral index of 
primordial scalar fluctuations) together with the assumed fiducial parameters for 
the $M-c$ relation.

\begin{table} 
\begin{center}
\begin{tabular}{|cccccc|}
\hline
\multicolumn{6}{|c|}{Cosmological parameters}\\
\multicolumn{6}{|c|}{WMAP9 + SNLS3 +BAO}\\
\hline
$\Omega_m$ & $\Omega_b$ & $\sigma_8$ & $H_0$ & $\tau$ & $n_s$ \\
\hline
$0.291$ & $0.047$ & $0.828$ & $68.98$ & $0.087$ & $0.969$ \\
\hline
\end{tabular}
\begin{tabular}{|ccc|}
\hline
\multicolumn{3}{|c|}{$M-c$ relation}\\
\multicolumn{3}{|c|}{parameters}\\
\hline
$c_0$ & $\alpha$ & $\sigma_{\ln c}$ \\
\hline
$4$ & $-0.1$ & $0.15$ \\
\hline
\end{tabular}
\caption{Fiducial cosmological and $M-c$ relation parameters.}
\label{par}
\end{center}
\end{table}

\subsection{$M_{ap}$ probability distribution function}\label{PDF}

A significant part of the noise contribution to $M_{ap}$ comes from 
intrinsic ellipticity and the finite number of background galaxies 
used to measure the shear signal. Nevertheless, 
the accuracy with which the lensing signal 
of individual halos can be measured is also limited by the presence 
of large--scale structures (LSS) along the line--of--sight,  
representing an additional statistical 
source of noise as discussed in \cite{Hoekstra01,maturi05} 
(see also \cite{SWJK}).

Here we neglect the effects on $M_{ap}$ measurements due to intrinsic 
alignments of background galaxies and LSS projections 
correlated with halo lens (although they might introduce non--negligible bias
and noise as discussed in \cite{marian09,gruen}) and we 
mainly focus on the impact of the concentration scatter and of 
the statistical noise from uncorrelated intrinsic ellipticities 
and uncorrelated LSS.
 
Following \cite{maturi05} we roughly split the total signal in two 
independent contributions: the signal from nonlinear scales due exclusively to
massive halos and the LSS signal generated by linearly evolved matter 
perturbations. Although questionable for many WL applications, 
it would be sufficient for our purposes. 
The observed $M_{ap}$ is then given by the sum of halo and LSS signals plus 
the noise contribution $M_{ap}^{g}$ due to intrinsic ellipticity 
of the galaxies and survey shot noise:
\begin{equation}
 M_{ap} = M_{ap}^{halo} + M_{ap}^{LSS} + M_{ap}^{g}
\label{obsmap}
\end{equation}

Under the assumption of uncorrelated intrinsic ellipticities 
and uncorrelated LSS both galaxy and LSS noise fields can be modeled as
isotropic Gaussian random fields with zero mean, 
$\left< M_{ap}^{g} \right> =\left< M_{ap}^{LSS} \right> = 0$, 
and variances given respectively by (see \cite{SWJK,Hoekstra01,maturi05} for
the derivation):
\begin{equation}
 \sigma^2_{g}(\vartheta_0) = \frac{\pi \sigma_\epsilon^2}{n} 
\int_0^{\vartheta_0} d\vartheta \vartheta Q^2(\vartheta) 
\end{equation}
($\sigma_\epsilon$ is the ellipticity dispersion of sources, $n$ 
is the average number density of galaxies inside the aperture), and:

\begin{equation}\label{m_lss}
\sigma_{LSS}^{2}(\vartheta_0) = 2\pi \int_{0}^{\infty} dl ~ l P_{\kappa}(l)g^2(l,\vartheta_0)  
\end{equation}
Here, the function $g(l,\vartheta_0)$ reads:
\begin{equation}
 g(l,\vartheta_0) = \int_{0}^{\vartheta_0} d\vartheta ~\vartheta U(\vartheta) 
J_0(l\vartheta)
\end{equation}
while the power spectrum  of the LSS convergence $P_{\kappa}(l)$
is related to the power spectrum of the three-dimensional density 
fluctuations $P_{\delta}(k)$ by Limber's equation: 
\begin{equation}
  P_{\kappa} (l)= \frac{9 H_{0}^2 \Omega_{m}^2}{4 c^2} \int_{0}^{\chi_h} d\chi
 \frac{{W}^{2}(\chi)}{a^2(\chi)} P_{\delta} \left( \frac{l}{D(\chi)},
\chi \right)
\end{equation}
($\chi$ and $D(\chi)$ being the comoving distance and the comoving 
angular diameter distance respectively). 
For a given distribution of source galaxies, $p(\chi)d\chi=p(z)dz$, 
the weight function $W$ reads:
\begin{equation}
W(\chi) = \int_{\chi}^{\chi_h} d\chi' p(\chi) \frac{D(\chi'-\chi)}{D(\chi')}
 \end{equation}
where $\chi_h$ is the distance to the last scattering surface.

The Gaussian noise $\sigma_{n}^2=\sigma_{g}^2+\sigma_{LSS}^2$ then
leads to a 
probability distribution function (PDF) for $M_{ap}$ given by:
\begin{equation}
 p_{n}(M_{ap}|M)=\frac{1}{\sqrt{2\pi}\sigma_{n}}
\exp \left[-\frac{1}{2} \left(\frac {M_{ap}- M_{ap}^{halo}(M)}{\sigma_{n}} \right)^2\right]
\label{pdfnoise}
\end{equation}
where $M_{ap}^{halo}(M)$ is the 
true value of the aperture mass of a halo with
mass $M$.

Although observations and simulation results 
tell us that, for a given mass $M$, halo concentrations are log--normally 
distributed with dispersion $\sigma_{\ln c}\sim 0.15-0.3$ \cite{Jing00}: 
\begin{equation}
p(c|M)= \frac{1}{c\sqrt{2\pi}\sigma_{\ln c}}
\exp \left[-\frac{1}{2} \left(\frac{\ln c-\ln \hat c(M)}{\sigma_{\ln c}} 
\right)^2\right]
\label{pconc}
\end{equation}
($\hat c(M)$ being the median concentration), it is common practice 
in theoretical calculations to use the PDF (\ref{pdfnoise}) where
$M_{ap}^{halo}$ is calculated assuming the median $M-c$ relation
and neglecting any dispersion (hereafter, we will drop the 
subscript $_{200}$ to denote the halo concentration). 

In order to take the concentration scatter into account, one should 
instead consider the convolution:
\begin{equation}
p(M_{ap}|M)=\int dM_{ap}^{halo} ~p_{n}(M_{ap}|M,c) ~p_{c}(M_{ap}^{halo}|M)
\label{pdf}
\end{equation}
where:
\begin{equation}
p_{c}(M_{ap}^{halo}|M)dM_{ap}^{halo} = p(c|M) dc
\label{pdfc}
\end{equation}
and now $M_{ap}^{halo}(M)=M_{ap}^{halo}(M,c)$.
It is then easy to show with a little algebra that the mean value of  
$M_{ap}$ is:
\begin{equation}
\left< M_{ap} \right> = \int dM_{ap} ~M_{ap} ~p(M_{ap}|M)= 
\int dM_{ap}^{halo} ~M_{ap}^{halo} ~p_{c}(M_{ap}^{halo}|M)=
\left< M_{ap}^{halo} \right>
\end{equation}
while the total $M_{ap}$ variance $\sigma$ reads:
\begin{eqnarray}
\nonumber
\sigma^2  &=& \int dM_{ap} \left(M_{ap}- \left< M_{ap} \right>\right)^2
~p(M_{ap}|M)  \\ \nonumber
 &=&  \int dM_{ap}^{halo}\left(M_{ap}^{halo}-\left< M_{ap}^{halo} \right>\right)^2 p_c(M_{ap}^{halo}|M) + \sigma_n^2 \\
&=& \sigma_c^2 + \sigma_n^2
\end{eqnarray}
Note that, because of the non--linearity of the $M-c$ relation $\left< M_{ap} \right> = \left< M_{ap}^{halo} \right> \neq M_{ap}^{halo}(\left< c \right>)$ and the same holds for the median value $\hat M_{ap}$.

Finally, we define the signal-to-noise ratio, {\it snr}, associated to each 
peak detection as: 
\begin{equation}
snr = \frac{\hat M_{ap}} {\sigma}
\label{sn}
\end{equation}
(here we have defined $snr$ in terms of the median $\hat M_{ap}$
since $p(M_{ap}|M)$ is expected to be non--symmetric because of the 
asymmetry of (\ref{pconc})).

\begin{figure}[]
\begin{center}
\includegraphics[angle=-90,scale=.30]{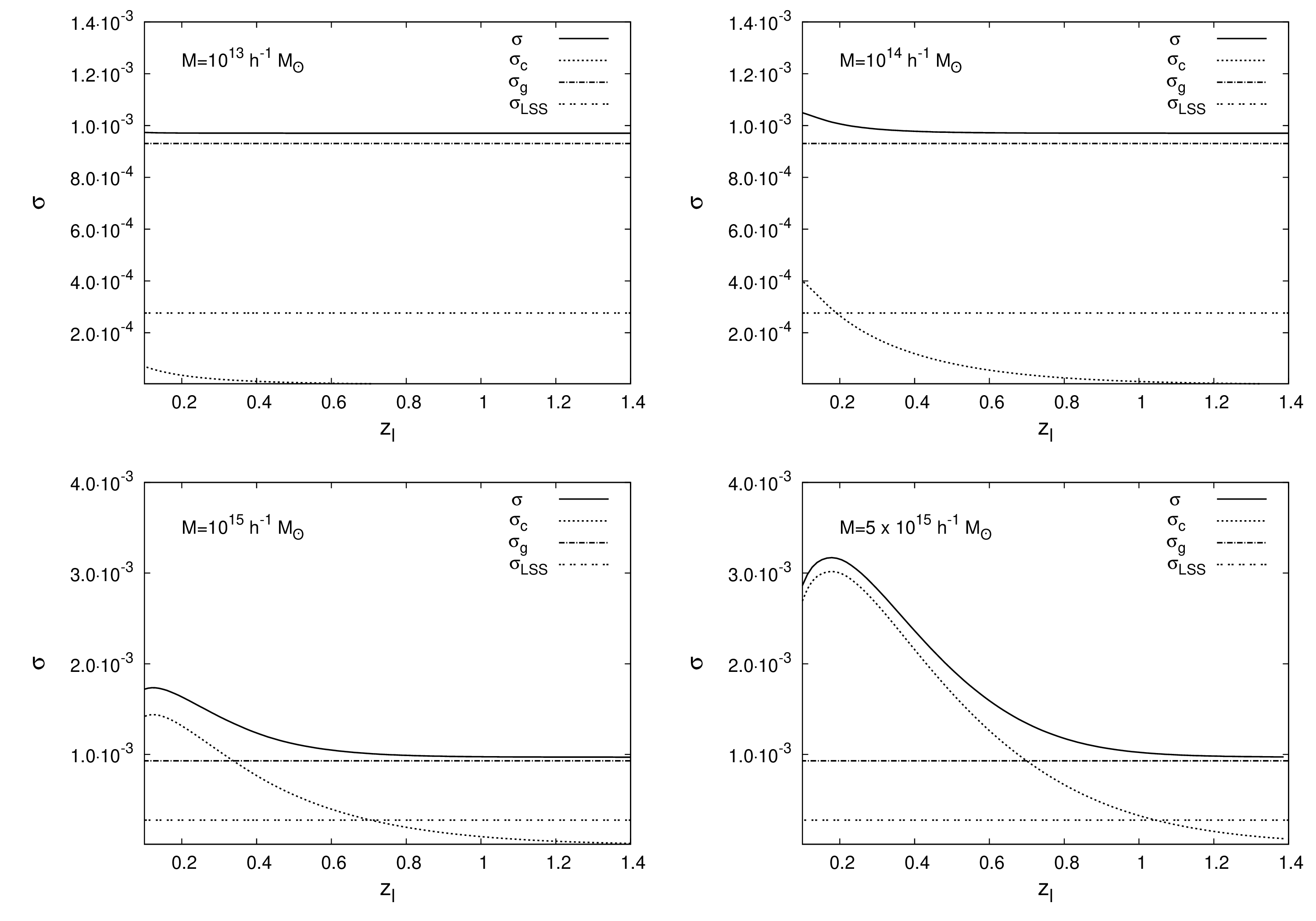}
\caption{Contributions to the total $M_{ap}$ variance $\sigma$ as function
of lens redshift and mass, due to intrinsic ellipticities and finite numer 
of background galaxies ($\sigma_g$), LSS ($\sigma_{LSS}$) and concentration 
scatter ($\sigma_c$).}     
\label{var}
\end{center}
\end{figure}

\begin{figure}[]
\begin{center}
\includegraphics[angle=-90,scale=.30]{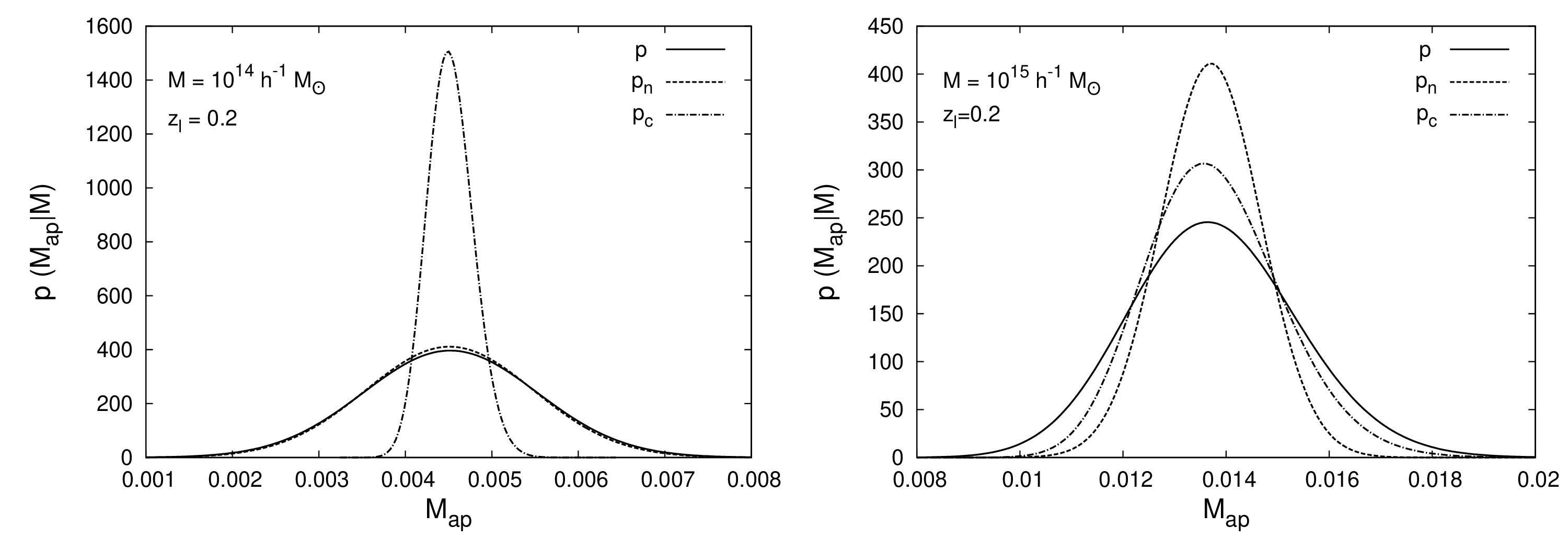}
\caption{PDF ($p(M_{ap}|M)$) for lens masses of $10^{14}$ (left panel) 
and $10^{15} h^{-1}M_{\odot}$ (right panel) at $z_h=0.2$ compared to the PDFs in 
the cases when only the statistical noise is considered ($p_n(M_{ap}|M)$), 
or only the effect of $M-c$ relation scatter is taken into 
account ($p_c(M_{ap}|M)$).}
\label{pdfplot}
\end{center}
\begin{center}
\includegraphics[angle=-90,scale=.20]{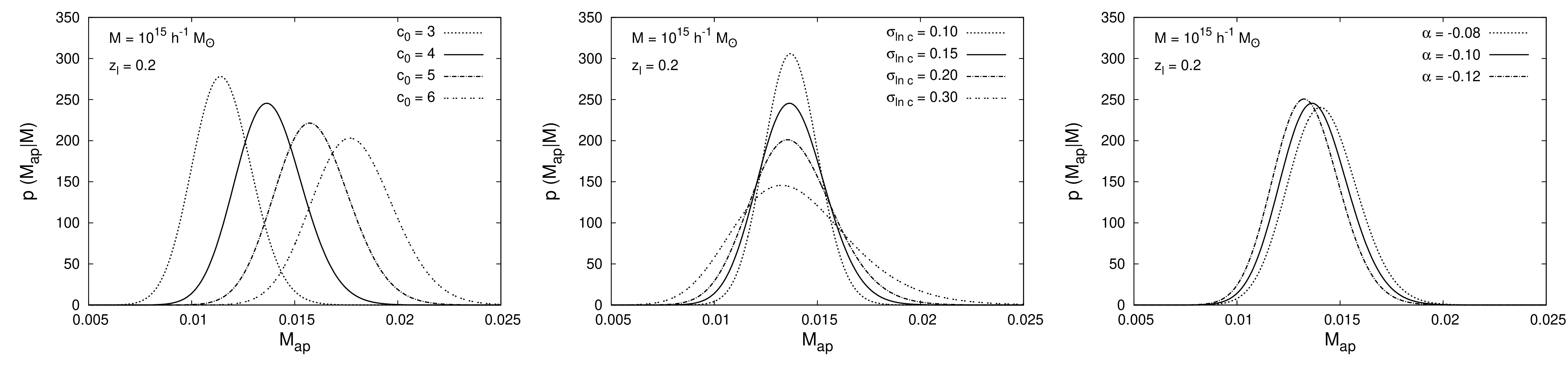}
\caption{Effect of varying the $M-c$ relation parameters, 
$c_0$ (left panel), $\sigma_{\ln c}$ (central panel) 
and $\alpha$ (right panel), on $M_{ap}$ PDF. Results are shown for an halo of 
mass $M = 10^{14} h^{-1} M_{\odot}$ located at $z_h=0.2$.}
\label{pdfMc}
\end{center}
\end{figure}

It is worth noticing that, unlike the statistical noise $\sigma_n$ which is not
related to  halo properties, $\sigma_c$ depends on both lens mass 
and redshift $z_h$, it increases (decreases) with $M$ ($z_h$).  
This is shown in fig. \ref{var} 
where the different contributions to the total variance $\sigma$
are displayed as a function of $z_h$ for different halo masses in the range
$10^{13}-5\cdot10^{15} ~ h^{-1}M_{\odot}$. Inspection of the figure indicates
that the scatter in $M-c$ relation affects significantly the $M_{ap}$ 
variance (and hence $snr$) of
halos with higher mass at relatively low redshifts. Its impact, however, 
becomes negligible at higher $z_h$. 
Fig. \ref{pdfplot} shows the PDF $p(M_{ap}|M)$ obtained from (\ref{pdf}) 
for lens masses of $10^{14}$ (left panel) and $10^{15} h^{-1}M_{\odot}$ 
(right panel) at $z_h=0.2$.
For comparison we also show the PDFs in the cases when only the statistical 
noise is considered ($p_n(M_{ap}|M)$, eq. (\ref{pdfnoise})) or only the effect 
of $M-c$ relation scatter is taken into 
account  ($p_c(M_{ap}|M)$, eq. (\ref{pdfc})). 
In the former case, $\sigma_c \simeq \sigma_{LSS} \ll \sigma_g $
so that $p \simeq p_n$ while in the latter $\sigma_c$ and $\sigma_g$ are of 
the same order of magnitude (see also fig. \ref{var}). 

The distorsions
in the PDF caused by variations in the parameters $c_0$, $\sigma_{\ln c}$ and 
$\alpha$ of the $M-c$ relation are shown in the left, central and right panels 
of fig. \ref{pdfMc}. The solid line indicates the fiducial model prediction.
Increasing the normalization $c_0$ shifts the mean value 
$\left< M_{ap} \right>$ to larger values while changes in $\sigma_{\ln c}$ mainly
affect the width and the skewness of the distribution which is only 
slightly altered by variations in $\alpha$.  

\begin{figure}[]
\begin{center}
\includegraphics[angle=-90,scale=.20]{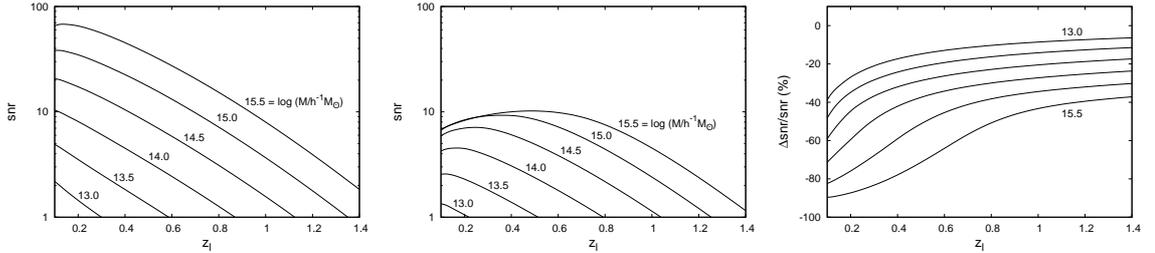}
\caption{Dependence of $snr$ on lens mass and redshift $z_h$. In the left panel
only statistical noise is considered while in central panel the concentration 
scatter is also taken into account. The relative deviation of $snr$ between 
the two cases is shown in the right panel.
Results are given for masses in the range
$10^{13}-5 \cdot 10^{15} h^{-1} M_{\odot}$ in steps of $\Delta \log M =0.5$.}
\label{snzM}
\end{center}
\end{figure}

The dependence of $snr$ on lens mass and redshift is shown in fig. \ref{snzM}.
In the left panel only statistical noise is considered while in the 
central panel we also include the concentration scatter. The right panel
displays the relative deviation $\Delta snr/snr$ between them. As discussed 
above, concentration scatter significantly increases the $M_{ap}$ variance of  
more massive halos. In turn, the associated $snr$ is strongly suppressed to no
more than $snr=10$.  

We have also checked that shifts in the cosmological parameters $\sigma_8$, 
$\Omega_m$, $H_0$ and $n_s$ do not appreciably modify the 
$M_{ap}$ PDF and $snr$.
This fact that they are almost 
independent of the underlying cosmology, can be understood looking
at how cosmology enters in the evaluation of $M_{ap}$ and $\sigma$. 
The convergence $\kappa(\vartheta)$ depends on the integrated Hubble 
rate through the lensing efficiency function $D_h D_{hg}/D_g$ 
which only weakly 
depends on cosmology since it involves a ratio of distances which is then 
further integrated over the source redshift. 
Further, $M_{ap}$ is independent of $n_s$ and $\sigma_8$. 
On the other hand, $n_s$ and $\sigma_8$ affect the LSS 
convergence power spectrum $P_\kappa(l)$ entering the total noise 
$\sigma$ through the $\sigma_{LSS}$ term. Nevertheless, 
this latter is always about one order of magnitude or more smaller than 
$\sigma_g$, yielding a negligible effect on $snr$. 

\subsection{Halo mass function}

Different semianalytic approaches and fits to simulation results,
aimed to model the abundance of halos 
\cite{ps,st,jenk,reed,warren},
can be characterized by the scaled differential halo mass function 
introduced in \cite{jenk}:
\begin{equation}
f(\sigma_M,z)\equiv \frac{M}{\rho_m(z)} 
\frac {dn_{halo}(M,z)}{d \ln \sigma^{-1}_M}
\label{mf}
\end{equation}
where $n_{halo}(M,z)$ is the number density of halos with mass $M$, $\rho_m$ is
the background matter density and $\sigma_M=\sigma_M(M,z)$ is the variance 
of the linear density field on the scale $M$. This definition of mass function has the advantage
that  to a good accuracy it does not explicity depend on redshift, matter 
power spectrum or cosmological parameters, their dependence being all enclosed 
in $\sigma_M$. Hence, this {\it ``universality''} of the mass function suggests 
that the form of $f$ should hold in all Gaussian hierarchical clustering model,
at all times and in any cosmology. Indeed, this has been verified to a good 
accuracy in several works by means of N--body simulations for a vast class of 
cosmologies (see, however, \cite{courtin} for a recent discussion about 
the topic of the universality of the mass function).

For its simplicity, in the following, we make use of the fitting form provided 
in \cite{jenk}:
\begin{equation}
f(\sigma_M)= 0.315 ~\exp \left[-|\ln \sigma_M^{-1}+0.61|^{3.8}\right]
\label{jen}
\end{equation}

\subsection{Peak number counts}

Having determined the theoretical mass function and having detailed 
how the $M_{ap}$ PDF  and the $snr$ for 
peak detection can be computed, we can now 
estimate the abundance of halos that produce significant 
peaks in the aperture mass map.  

Noticing that the probability for a halo of mass $M$ to be detected as a 
shear peak with $snr$ above the threshold $snr^*=\hat M^*_{ap}/\sigma$ is:
\begin{equation}
P(>snr^*|M)=\int_{M^*_{ap}} dM_{ap} ~p(M_{ap}|M)
\end{equation}
the shear peak function expected to be observed then reads:  
\begin{equation}
N_{halo}(>snr^*)=\int dV \int dM 
\frac {dn_{halo}(M,z_h)}{dM} P(>snr^*|M)
\label{nobs}
\end{equation}
where $dV$ is the comoving volume element per unit solid angle\footnote{$dV=
d\chi ~\chi^2  = dz \left[\int_0^z dz'/H(z')\right]^2 
d\chi/dz$ in spatially 
flat cosmologies, $\chi$ being the comoving distance.}.
The above expression gives the observed number of halo peaks  
taking into account, through (\ref{pdf}), the scatter in $M_{ap}$ due to 
statistical noise and concentration distribution.

\begin{figure}[]
\begin{center}
\includegraphics[angle=-90,scale=.30]{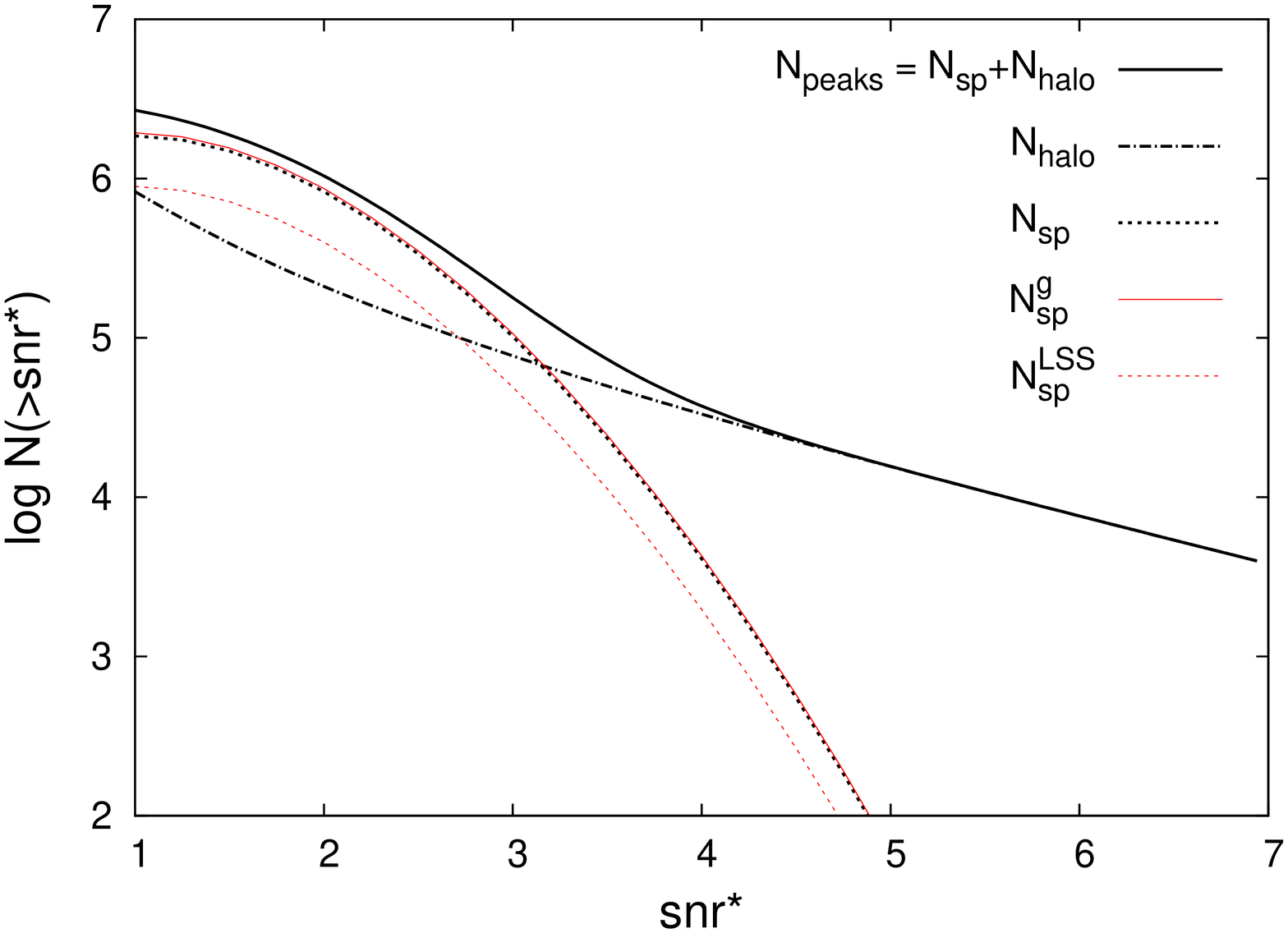}
\caption{Number of detectable peaks as a function of $snr^*$ for the 
fiducial model. Black lines refer to 
total number (solid) and the contribution from halo ( dash--dotted) and 
spurious (dotted) peaks.
Red lines refer to spurious peaks generated by galaxy noise (solid) 
and LSS (dotted).}  
\label{ntot}
\end{center}
\begin{center}
\includegraphics[angle=-90,scale=.30]{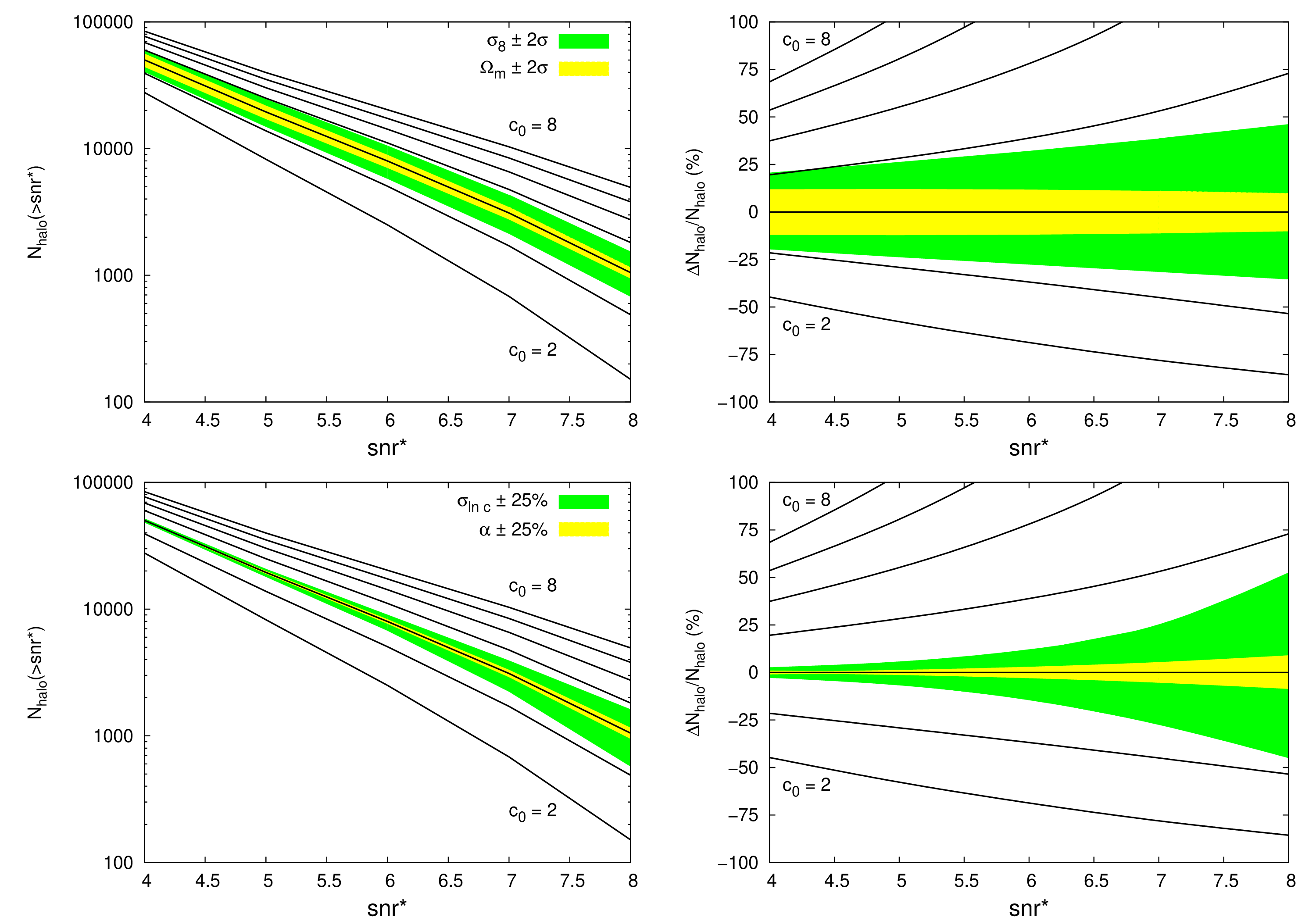}
\caption{{\it Left panels}: 
Total number of halo peaks as a function of $snr^*$ in the redshift range 
$\Delta z=0.1-1.4$. Shaded regions indicate deviations from the fiducial 
model predictions arising from shifts in $\sigma_8$ and $\Omega_m$ within the
current $2-\sigma$ bounds (top panels), and $\pm 25$\% offsets in $\alpha$ and 
$\sigma_{\ln c}$ (bottom panels). Solid lines refer to changes in $c_0$   
(from $2$ to $8$ in steps of $1$ from the lowest to highest curve). 
{\it Right panels}: Relative deviation $\Delta N_{halo}/N_{halo}=
[N_{halo}-N_{halo,fid}]/N_{halo,fid}$.}
\label{nsn}
\end{center}
\end{figure}
\begin{figure}[]
\begin{center}
\includegraphics[angle=-90,scale=.30]{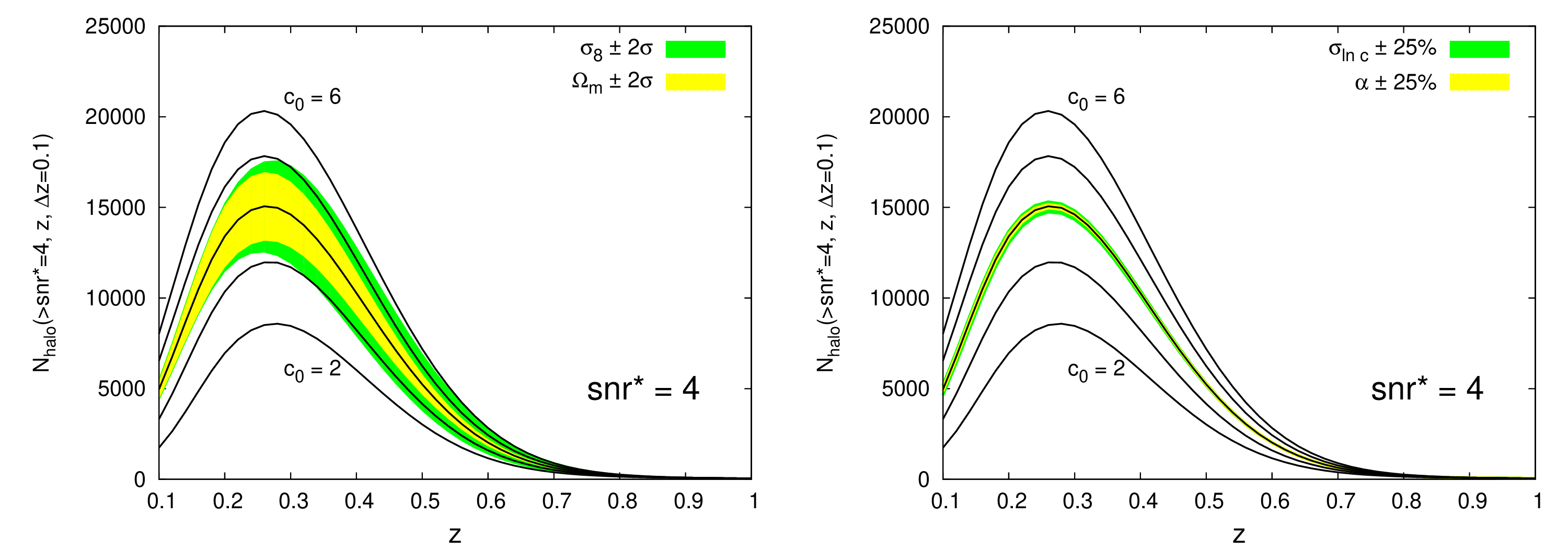}
\includegraphics[angle=-90,scale=.30]{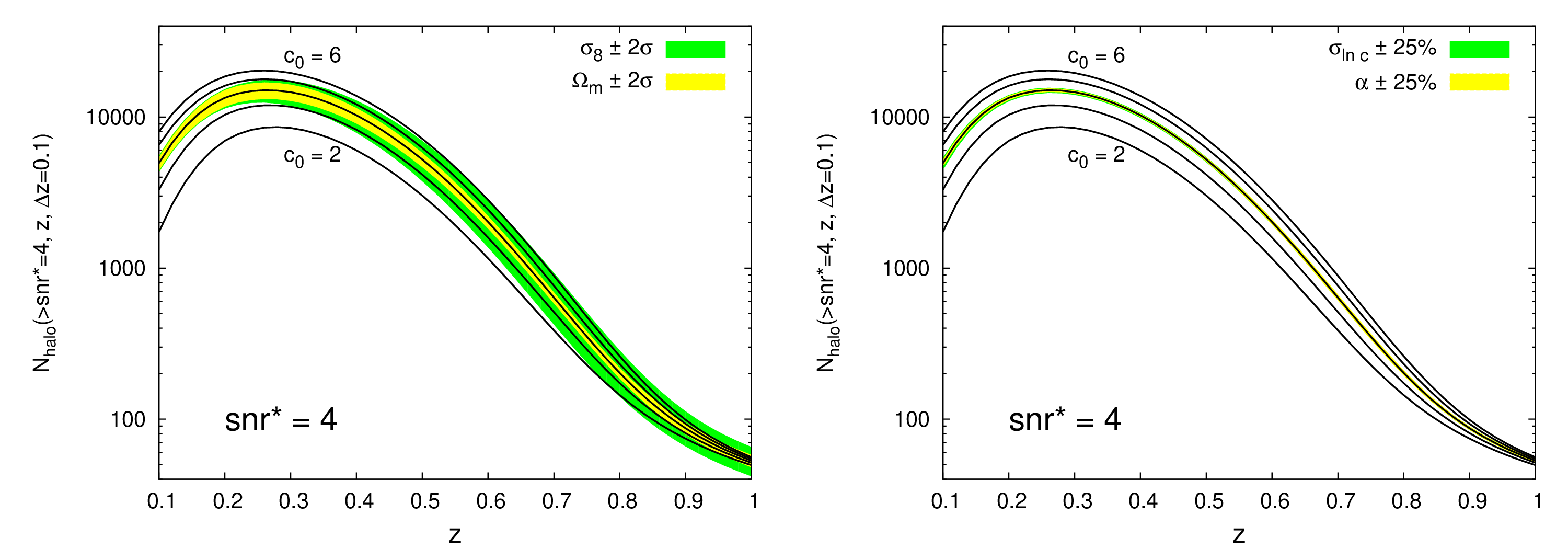}
\caption{{\it Top panels}: Number of halo peaks in redshift bin of 
$\Delta z = 0.1$ as a function
of $z$ for $snr^*=4$. Shaded regions indicate deviations from the fiducial 
model predictions arising from shifts in $\sigma_8$ and $\Omega_m$ within the
current $2-\sigma$ bounds (left panel), and $\pm 25$\% offsets in $\alpha$ and 
$\sigma_{\ln c}$ (right panel). Solid lines refer to changes in $c_0$   
(from $2$ to $6$ in steps of $1$ from the lowest to highest curve). 
{\it Bottom panels}: Same as top panels but plotted in logarithmic scale 
to emphasize the behavior at high redshifts.}
\label{nz}
\end{center}
\end{figure}
\begin{figure}[]
\begin{center}
\includegraphics[angle=-90,scale=.30]{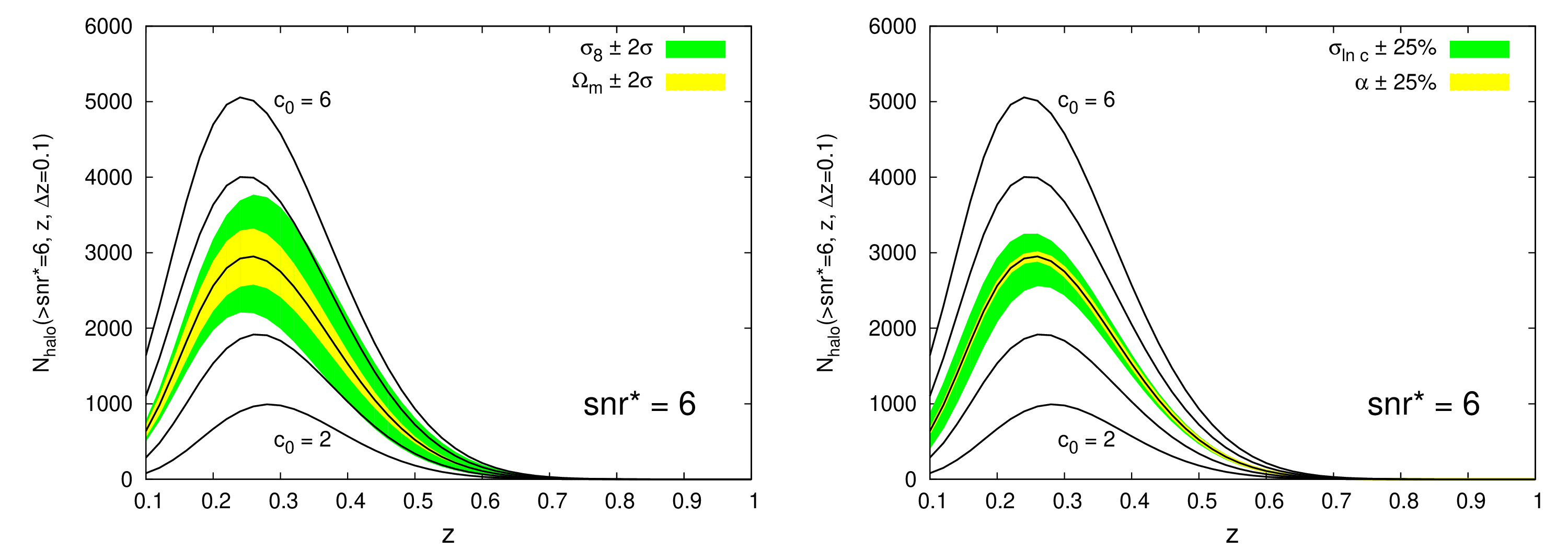}
\includegraphics[angle=-90,scale=.30]{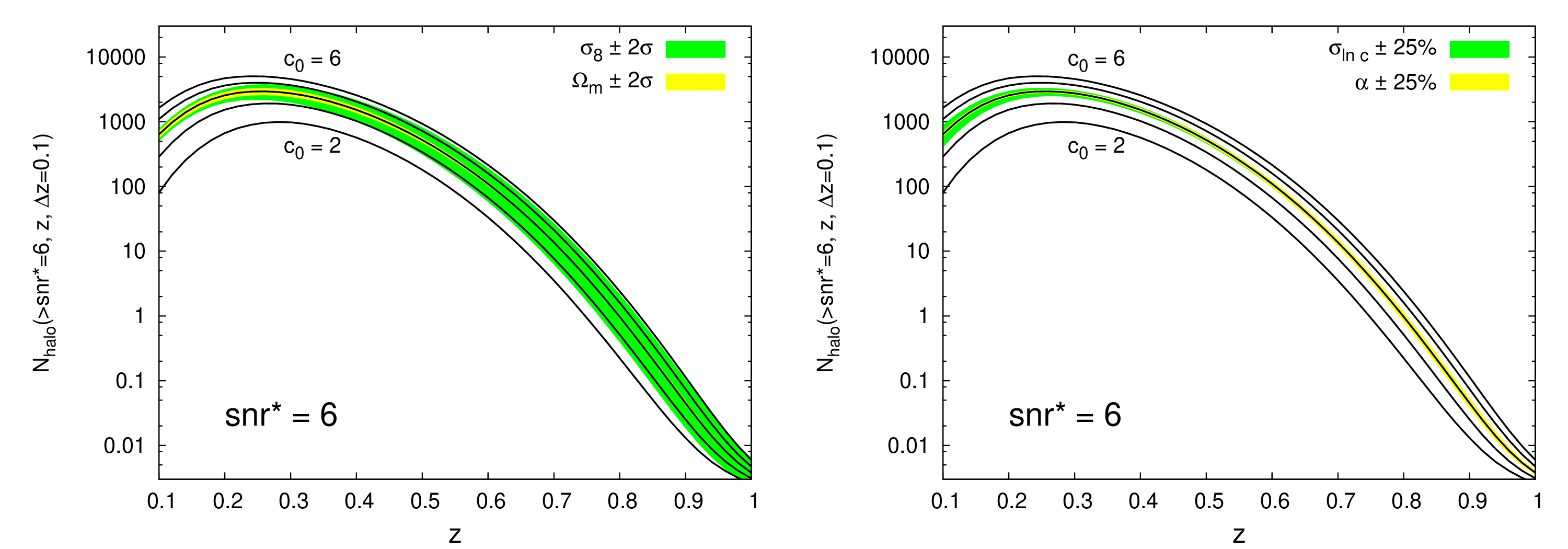}
\caption{As in previous figure but for $snr^*=6$.}
\label{nzsn6}
\end{center}
\end{figure}

The total number of observed shear peaks is, however, 
larger than $N_{halo}$ because
of spurious peaks caused by projection effects of LSS as well as shape 
and shot noise of background galaxies. 

An analytic approach suitable to quantify the level
of spurious detections, has been developed and applied in \cite{maturi10}. 
Analytic results were there compared to peak counts from numerical ray--tracing 
simulations finding good agreement for different filter 
functions.
We remind here that this approach relies 
on peak counts in random Gaussian fields \cite{bardeen},
very well representing noise and LSS but not non--linear objects 
such as virialized halos which are highly non--Gaussian. 
We also remind that, following 
\cite{maturi05} and according to eq. (\ref{obsmap}), 
we define the LSS signal as being due to linearly evolved structures alone 
so that a Gaussian description applies (see section \ref{PDF}).

According to \cite{maturi10}, the number of spurious peaks may be estimated as:
\begin{equation}
N_{sp}(>snr^*)=\frac{1}{(2\pi)^{\frac{3}{2}}} \left(\frac{\sigma_1}{\sigma_n}\right)^2 snr^* \exp \left[ - \frac{snr^{* 2}}{2} \right]
\label{spurious}
\end{equation}
where:
\begin{equation}
\sigma_1^2=\int dl ~l^3 P(l) \left|\hat Q(\vartheta) \right|^2
\end{equation}
Here $\hat Q(\vartheta)$ is the Fourier transform of the filter while 
$P(l)=P_n+P_\gamma(l)$ is the power spectrum of the statistical noise 
($P_n=\sigma_\epsilon^2/2n_{gal}$ and
$P_\gamma(l)=P_{\kappa}(l)/2$ being the power spectra of galaxy noise and LSS 
shear respectively).

The total number of observed peaks is finally given by:
$$
N_{peaks}(>snr^*)=N_{halo}(>snr^*)+N_{sp}(>snr^*)
$$
Fig. \ref{ntot} shows $N_{peaks}$, $N_{halo}$ and $N_{sp}$ as a function of 
the threshold $snr^*$ for our fiducial model.
Spurious peaks generated by galaxy noise alone $N_{sp}^g$ and
LSS alone $N_{sp}^{LSS}$ are also shown for comparison.

It is worth noticing that the abundance of spurious detections (although
it can vary with the filter and/or the filter scale, see \cite{maturi10})
is almost independent of the underlying cosmology, 
the only dependence on cosmology being ascribed to
the LSS power spectrum $P_{\kappa}(l)$ entering the ratio 
$\sigma_1 /\sigma_n$. 
Indeed, as observed at the end of section \ref{PDF} (see also fig. \ref{var}), 
the LSS contribution to $\sigma_n^2 = \sigma_g^2+\sigma_{LSS}^2$ is always about
one order of magnitude smaller than the galaxy noise contribution and the 
same holds for  $\sigma_1^2 = \sigma_{1,g}^2+\sigma_{1,{LSS}}^2$.
Hence, $N_{sp} \simeq N_{sp}^g$ (see fig. \ref{ntot}) and the cosmological 
dependence is strongly suppressed. 
We have checked that no substantial changes in the number of spurious peaks
occur when varying the cosmological parameters within 
the $2-\sigma$ current bounds.

As observed in \cite{hamana04,maturi10} 
spurious peaks dominates the counts at relatively 
small peak amplitudes while the 
halo contribution becomes relevant slightly above $snr^* \sim 3$ and dominates 
at higher values. Since we are mainly interested in how peak counts can 
constrain the $M-c$ relation, we have to select those peaks associated 
to real halos.
Thus, in the following, 
we consider only $snr^* \geq 4$ (otherwise, we implicitly assume that spurious 
contaminations are avoided by means of suitable techniques, 
see below).

In the previous Section we have shown how the $M_{ap}$ PDF and $snr$ strongly 
depend on the $M-c$ relation while it is only scarcely affected by 
variations in the cosmological parameters. 
Halo concentrations are then anticipated to be crucial in 
determining the number of halo peaks.
Nevertheless, $\sigma_8$ and $\Omega_m$ are intimately related to the halo 
mass function, eq. (\ref{mf}), and hence expected to largely affect
the peak function as well. 

We  then investigate the impact of halo concentration and 
cosmological parameters $\sigma_8$ and $\Omega_m$ on weak lensing peak counts.
We do not consider the effects of $H_0$ and $n_s$ which have only a 
minor impact on the halo mass function. 

Deviations from reference model predictions 
arising from shifts in normalization $c_0$, slope $\alpha$ and scatter 
$\sigma_{\ln c}$ of $M-c$ relation as well as 
cosmological parameters $\Omega_m$ and $\sigma_8$ are considered and 
compared in Fig. \ref{nsn}.
The total number of halo peaks $N_{halo}(>snr^*)$ in the redshift range 
$\Delta z= 0.1-1.4$ is shown in the left panels as a function of $snr^*$ while
the right ones display the relative deviation, 
$\Delta N_{halo}/N_{halo}=[N_{halo}-N_{halo,fid}]/N_{halo,fid}$,
from the prediction of the fiducial model. Note that, although 
future galaxy surveys
will likely extend over a much greater redshift range, as it will 
be shown later the number of halo peaks is negligible for $z \gtrsim 0.9$.
Shaded regions in figure indicate departures in cosmological 
parameters (top panels) 
within the current $2-\sigma$ bounds, and $\pm 25$\% offsets in
$\alpha$ and $\sigma_{\ln c}$ (bottom panels). 
Solids lines refers to changes in $c_0$.

Among $M-c$ relation parameters, the normalization $c_0$
has the major impact on peak counts,
yielding deviations from $\sim 25\%$ to $\sim 75\%$ at 
the increasing of $snr^*$ in the range considered, when changing $c_0$ 
of $\pm 1$. On the other hand, the effect of varying 
$\sigma_{\ln c}$ and $\alpha$ is quite modest (or negligible) at  
low $snr^*$. Nevertheless, the impact of $\sigma_{\ln c}$ rapidly increases 
at higher  $snr^*$.

Observing that the range within which $\Omega_m$ and $\sigma_8$ 
are varied correspond to deviations of only 
$\sim 3-4$\%, halo counts are certainly more sensitive to cosmological 
parameters.
Within this range, $\sigma_8$ yields deviations in halo number 
of $\sim 25- 50\%$ while they are limited to $\sim 10-15 \%$ when changes 
in $\Omega_m$ are considered.

\begin{figure}[]
\begin{center}
\includegraphics[angle=-90,scale=.30]{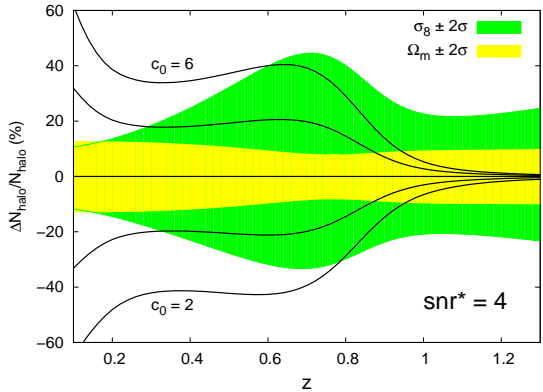}
\caption{Percentage deviation from the fiducial model peak function
due to shifts in $\sigma_8$, $\Omega_m$ and $c_0$ as in the left panel of fig.
\ref{nz}.}

\label{nzdev}
\end{center}
\end{figure}
\begin{figure}[]
\begin{center}
\includegraphics[angle=-90,scale=.30]{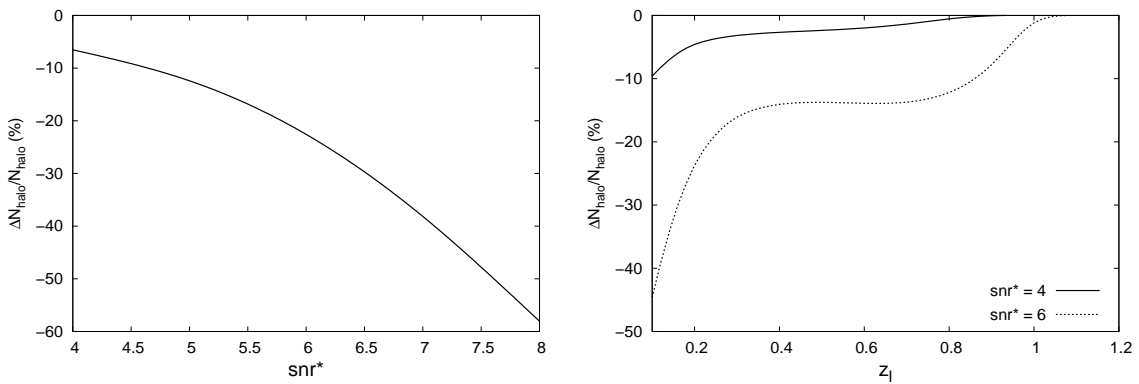}
\caption{Number counts are shifted toward lower values when
concentration scatter is taken into account. 
The relative shift in $N_{halo}(>snr^*)$ and 
$N_{halo}(>snr^*,z,\Delta z)$ from the case when 
only statistical noise is considered, is shown in the left and right panel
respectively.}
\label{nrap}
\end{center}
\end{figure}

So far, we considered the total number of halo peaks evaluated from 
(\ref{nobs}) by integrating over the full redshift range $\Delta z=0.1-1.4$
thus smoothing out the dependence of the halo mass function (\ref{mf})
and the $M_{ap}$ PDF on $z$. 
Actually, the evolution of the mass function depends on the 
underlying cosmology through $\sigma_8$ and $\Omega_m$, while the redshift 
dependence of the PDF is sensitive to the $M-c$ relation. Thus, it is 
worth to investigate how peaks are distributed in redshift bins in order
to get further insights on the parameters under investigation.  
To this aim we integrate (\ref{nobs}) over bins of $\Delta z=0.1$.

Note that, when dealing with real observations, redshift measurements 
are needed in order to assign a given halo peak to a bin. 
Redshift information can not be inferred from WL data alone, 
but additional data are required, 
e.g. optical data (see the discussion in section \ref{concl}).
On the other hand, spurious peaks are fake detections,
not related to a particular halo and not having any real optical counterpart
so that no redshift can be attributed to them. 
Here, we are implicitly and naively assuming that redshifts 
were properly assigned to halo peaks and spurious detections eliminated
from the sample, e.g. by keeping only sufficiently high peaks and/or
by matching WL and optical data (or by means 
of tomographic techniques, see e.g. \cite{HS05}).

The resulting peak redshift distribution $N_{halo}(>snr^*,z,\Delta z)$ is 
plotted in figs. \ref{nz} and \ref{nzsn6} for $snr^*=4$ and $6$ respectively
in both linear (top panels) and logarithmic 
(bottom panels) scale in the y--axis to emphasize the 
behavior at low or high redshifts.
Again, deviations from fiducial model predictions due to cosmological 
(left panels) and $M-c$ relation (right panels) parameters are shown.

Fig. \ref{nzdev} displays
the relative deviation, $\Delta N_{halo}/N_{halo}$ for the case $snr^*=4$ due 
to shifts in $c_0$, $\sigma_8$ and $\Omega_m$. Similar results are obtained
for $snr^*=6$ but with larger deviations. Inspection of the figure indicates
that probing the high--$z$ tail of the peak function, where it is most sensitive
to cosmological parameters, could therefore be an indirect method to 
investigate the halo mass function.
 
Finally, we want to show how the number counts change when the 
concentration scatter is taken into account. 
As already observed, scatter in $M-c$ relation can affect significantly 
the $M_{ap}$ variance reducing the $snr$ for halo peak detections.
The net effect is to shift the number counts toward lower 
values. The relative shift in $N_{halo}(>snr^*)$ and 
$N_{halo}(>snr^*,z,\Delta z)$ from the case when only statistical noise 
is considered, is displayed in the left and right panel, respectively, of 
fig. \ref{nrap}.

\section{Fisher matrix forecasts}
\label{fisher}

In this Section, we employ the Fisher matrix (FM) approach 
to probe the sensitivities of 
WL halo number counts to $M-c$ relation and cosmological
parameters ${\bf p}=(c_0,\alpha,\sigma_{\ln c},\sigma_8,\Omega_m)$.

\begin{table} 
\begin{center}
\begin{tabular}{|cccccc|c|}
\hline
$snr^*$ & $\sigma(c_0)$ & $\sigma(\alpha)$ & $\sigma(\sigma_{\ln c})$ & 
$\sigma(\sigma_8)$ & $\sigma(\Omega_m)$& priors\\
\hline
\multirow{2}{*}{$4$} & $0.598$ & $0.198$ & $0.128$ & $0.025$ & $0.027$ &\multirow{4}{*}{no priors}\\
 & $(14.9\%)$ & $(198.2\%)$ & $(85.3\%)$ & $(2.9\%)$ & $(9.2\%)$ & \\
\cline{1-6}
\multirow{2}{*}{$6$} & $1.535$ & $0.976$ & $0.096$ & $0.178$ & $0.181$ &\\
 & $(38.3\%)$ & $(975.7\%)$ & $(64.0\%)$ & $(21.5\%)$ & $(62.1\%)$ & \\
\hline

\multirow{2}{*}{$4$} & $0.392$ & $0.062$ & $0.073$ & $0.009$ & $0.009$ &\multirow{4}{*}{$\sigma_p(\sigma_8)=\sigma_p(\Omega_m)=0.01$} \\
 & $(9.8\%)$ & $(62.5\%)$ & $(48.6\%)$ & $(1.0\%)$ & $(3.0\%)$ & \\
\cline{1-6}
\multirow{2}{*}{$6$} & $0.416$ & $0.056$ & $0.022$ & $0.010$ & $0.010$ &\\
 & $(10.4\%)$ & $(56.5\%)$ & $(15.0\%)$ & $(1.2\%)$ & $(3.4\%)$ & \\
\hline

\multirow{2}{*}{$4$} & $-$ & $-$ & $-$ & $0.0025$ & $0.0024$ &\multirow{4}{*}{$c_0, \alpha, \sigma_{\ln c}$ fixed to fiducial values} \\
 & $-$ & $-$ & $-$ & $(0.3\%)$ & $(0.8\%)$ & \\
\cline{1-6}
\multirow{2}{*}{$6$} & $-$ & $-$ & $-$ & $0.0044$ & $0.0064$ &\\
 & $-$ & $-$ & $-$ & $(0.5\%)$ & $(2.2\%)$ & \\
\hline
\end{tabular}
\caption{Fisher matrix forecasted $1-\sigma$ uncertainties  on $c_0$, $\alpha$, 
$\sigma_{\ln c}$, $\sigma_8$ and $\Omega_m$. Results are obtained for 
$snr^*=4$ and $6$  assuming: i) no priors, ii) a prior of $0.01$ on both 
$\sigma(\sigma_8)$ and $\sigma(\Omega_m)$, iii) a perfect knowledge of $M-c$ relation. 
The percentages in parentheses denote the relative error $\sigma(p_i)/p_i$.}
\label{t3}
\end{center}
\begin{center}
\begin{tabular}{|c|rrrrr|}
\hline
$snr^*=4$& $c_0$ & $\alpha$ & $\sigma_{\ln c}$ & $\sigma_8$ & $\Omega_m$ \\
\hline
$\Omega_m$ & $-0.52$ & $0.93$ & $-0.84$ & $-0.81$ & $1.00$ \\
$\sigma_8$ & $-0.07$ & $-0.94$ & $0.42$ & $1.00$ & $ $ \\
$\sigma_{\ln c}$ & $0.85$ & $-0.69$ & $1.00$ & $ $ & $ $ \\
$\alpha$ & $-0.24$ & $1.00$ & $ $ &  $ $ &  $ $ \\
$c_0$ & $1.00$ & $ $ & $ $ & $ $ & $ $ \\
\hline
\end{tabular}
\end{center}
\begin{center}
\begin{tabular}{|c|rrrrr|}
\hline
$snr^*=6$ & $c_0$ & $\alpha$ & $\sigma_{\ln c}$ & $\sigma_8$ & $\Omega_m$ \\
\hline
$\Omega_m$ & $-0.69$ & $0.99$ & $0.68$ & $-0.97$ & $1.00$ \\
$\sigma_8$ & $0.49$ & $-0.99$ & $-0.84$ & $1.00$ & $ $ \\
$\sigma_{\ln c}$ & $0.06$ & $0.75$ & $1.00$ & $ $ & $ $ \\
$\alpha$ & $-0.61$ & $1.00$ & $ $ &  $ $ &  $ $ \\
$c_0$ & $1.00$ & $ $ & $ $ & $ $ & $ $ \\
\hline
\end{tabular}
\caption{Correlation coefficients $\rho_{ij} = \sigma_{ij} /\sigma(p_i)\sigma(p_j)$ for $snr^*=4$ (top) and $snr^*=4$ (bottom).}
\label{t2}
\end{center}
\end{table}
\begin{table} 
\begin{center}
\begin{tabular}{|cccc|}
\hline
$snr^*$ & $M_{\textrm {piv}}/10^{14}h^{-1}M_\odot$ & $c_{\textrm {piv}}$ & $\sigma_{c_{\textrm {piv}}}$ \\
\hline
$4$ & $1.19$ & $3.93$ & $0.57$ $(14.5\%)$ \\
%\hline
$6$ & $1.27$ & $3.91$ & $1.19$ $(30.5\%)$ \\
\hline
\end{tabular}
\caption{$M_{\textrm {piv}}$, $c_{\textrm {piv}}$, and $\sigma_{c_{\textrm {piv}}}$ for 
$snr^*=4$ and $6$.}
\label{tpiv}
\end{center}
\end{table}

\begin{figure}
\begin{center}
\includegraphics[angle=-90,scale=.15]{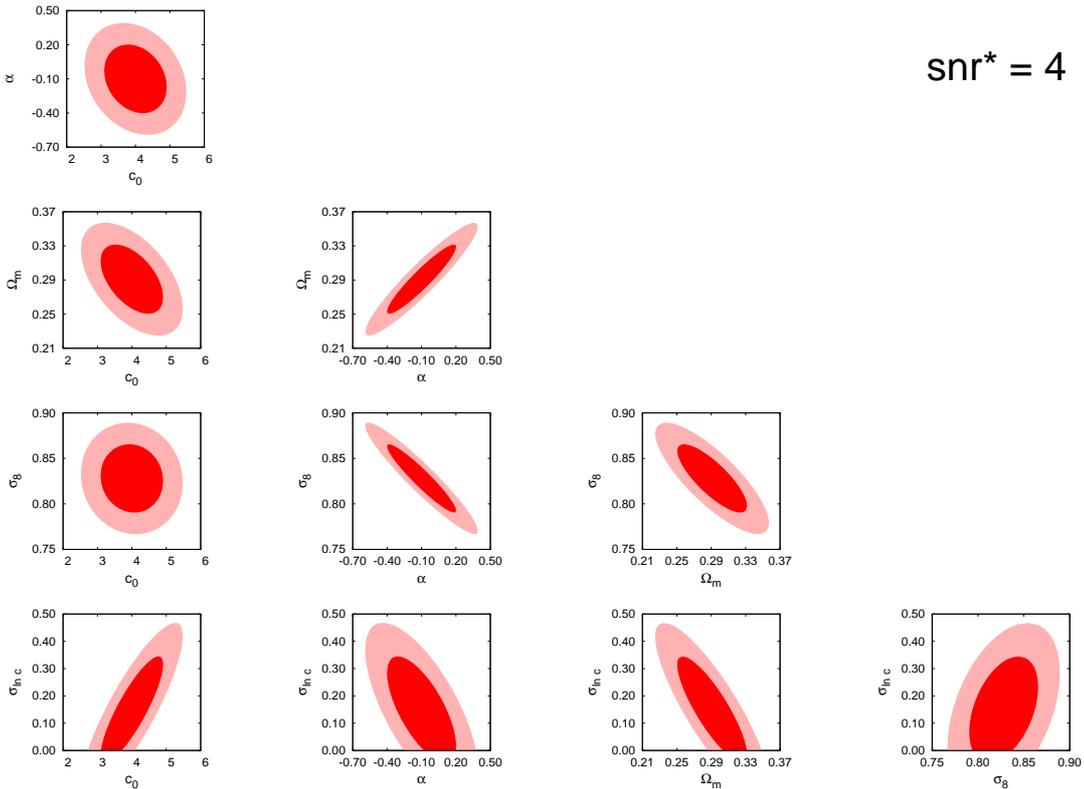}
\caption{$1-$ and $2-\sigma$ marginalized Fisher ellipses for $c_0$, $\alpha$, 
$\sigma_{\ln c}$, $\sigma_8$ and $\Omega_m$ in the case $snr^*=4$.} 
\label{fm}
\end{center}
\end{figure}

\begin{figure}
\begin{center}
\includegraphics[angle=-90,scale=.3]{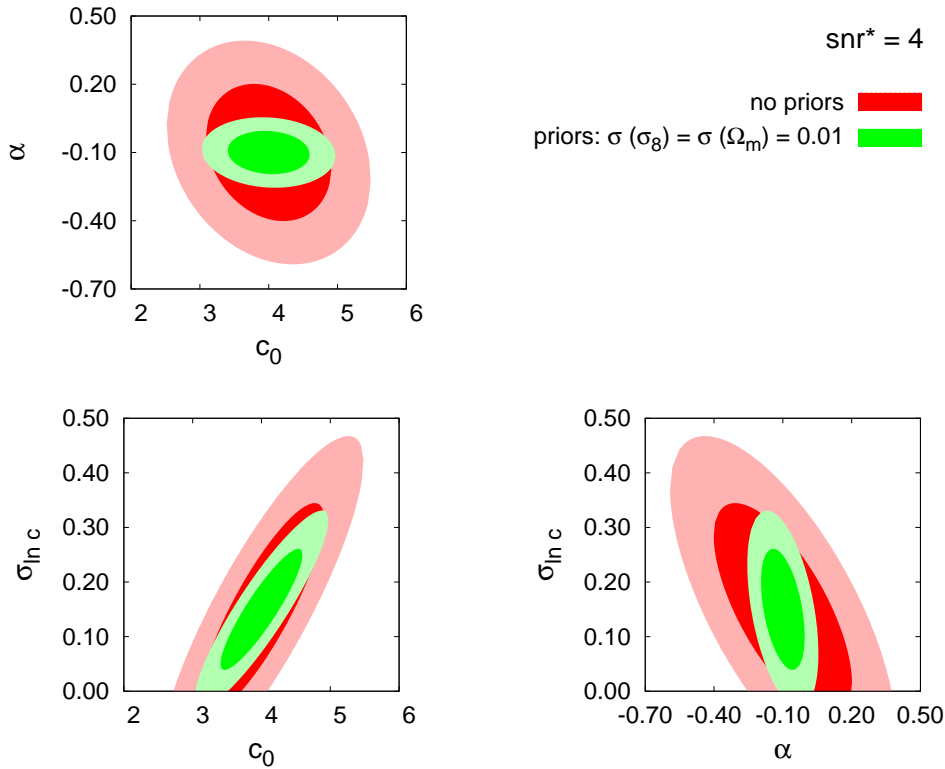}
\caption{$1-$ and $2-\sigma$ constraints on the $M-c$ relation parameters
($c_0$, $\alpha$, $\sigma_{\ln c}$) obtained for $snr^*=4$ assuming: 
i) no priors, 
ii) a prior of $0.01$ on both $\sigma(\sigma_8)$ and $\sigma(\Omega_m)$.}
\label{fmprior}
\end{center}
\end{figure}
\begin{figure}
\begin{center}
\includegraphics[angle=-90,scale=.3]{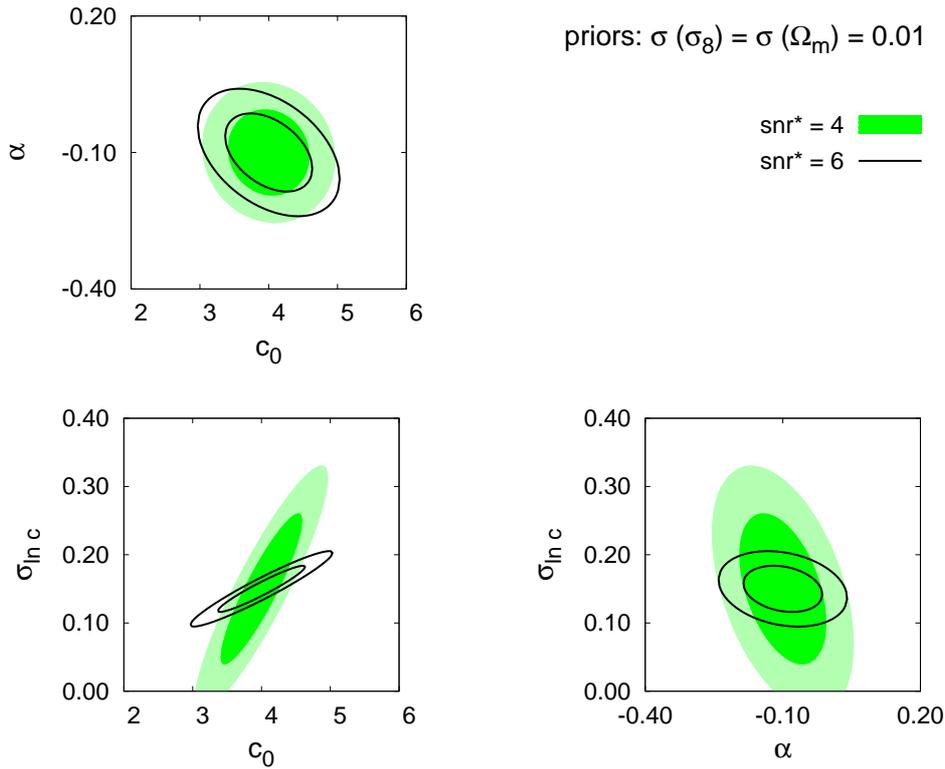}
\caption{$1-$ and $2-\sigma$ constraints on the $M-c$ relation parameters
($c_0$, $\alpha$, $\sigma_{\ln c}$). Results are obtained for $snr^*=4$ and $6$
placing a prior of $0.01$ on both $\sigma(\sigma_8)$ and $\sigma(\Omega_m)$.}
\label{fmprior2}
\end{center}
\end{figure}

Following \cite{holder}, the FM for the number of halo peaks 
$N_k(z_k,{\bf p})=N_{halo}(>snr^*,z_k,\Delta z$) in the $k$--th redshift bin 
of width $\Delta z=0.1$ can be written as:
\begin{equation}
F_{ij}=\sum^{n_{bin}}_{k=1}\frac{\partial N_k}{\partial p_i}
\frac{\partial N_k}{\partial p_j}\frac{1}{N_k^{fid}}
+ \frac {\delta_{ij}}{\sigma_p^2(p_i)}
\end{equation}
where $N_k^{fid}$ is the expected number of peaks in the fiducial model in the
$k$--th bin, $n_{bin}$ is the total number of bins and $\sigma_p(p_i)$
is a possible  Gaussian prior on the parameter $p_i$.
The inverse of FM 
then gives the covariance matrix $C_{ij}$, the diagonal elements of which represent
the lowest variance $\sigma(p_i)$ one can achieve on the parameter $p_i$.

The forecasted uncertainties $\sigma(p_i)$ turn out to depend on 
the threshold $snr^*$ and are summarized in table \ref{t3} together with the
percentage errors $\sigma(p_i)/p_i$ (in parentheses) for 
$snr^*=4$ and $6$. In addition, the correlation 
coefficients $\rho_{ij} = \sigma_{ij} /\sigma(p_i)\sigma(p_j)$ 
($\sigma_{ij}$ being the covariance between $p_i$ and $p_j$) are shown in 
table \ref{t2}.

We however warn the reader that the FM forecasts given in table \ref{t3},
should be taken as optimistic limits (see the discussion in the next Section).

In the absence of any prior information, we first note how peak statistics 
turns out to be competitive with other cosmological probes in placing 
constraints on $\sigma_8$ and $\Omega_m$. For $snr^*=4$ we indeed obtain 
$\sigma(\sigma_8)=0.025$ and $\sigma(\Omega_m)=0.027$ which are very 
close to the current limits from Cosmic Microwave Background (CMB) 
or LSS data.

Turning to $M-c$ relation, the most constrained parameter is the normalization 
$c_0$. For $snr^*=4$, we find $\sigma(c_0)/c_0 \sim 15 \%$, which,
at first glance, could look a not so encouraging result.
We stress, however, that this is still an appreciable result,
considered the wide variety of $c_0$ values reported in literature.
On the other hand, peak statistics alone seems less efficient in 
putting limits on 
$\sigma_{\ln c}$ and in particular on the slope $\alpha$ which is substantially 
unconstrained.
Going to $snr^*=6$ has the effect of reducing the overall number of peaks 
so that a worsening of the constraints is expected because of the 
poorer statistics. This is not, however,  the case for $\sigma_{\ln c}$.
This is likely related to the fact that larger $snr$ probe larger masses
which are more sensitive to $\sigma_{\ln c}$ especially at moderate
low redshifts where the number of halo peaks is still significative 
(compare figs. \ref{nz} and \ref{nzsn6}).

In light of these results, peak statistics alone seems not to be 
helpful in constraining the $M-c$ relation. 
We then consider the effect of combining peak counts with 
current/near--future observations in order to probe whether the constraints 
can be improved.
  
Current data from CMB, LSS and type Ia Supernovae already 
constrain $\Omega_m$ with  an uncertainty
$\sigma(\Omega_m) \sim 0.01$ while set the most stringent 
bounds on $\sigma_8$  to $\sigma(\sigma_8) \sim 0.015-0.018$, depending 
on the dataset used (see e.g \cite{wmap9,planck}; 
see also \cite{CFHT} for an analysis based on WL tomography).
Nevertheless, assuming an uncertainty of $\sigma(\sigma_8)=0.01$ 
is not unreasonable. 
It is just less than half of the current limit and, hopefully, it 
will be achieved by a joint analysis of current and/or near future 
cosmological probes (see, however, \cite{burenin} where cluster mass function 
measurements are used in combination with other dataset to give 
$\sigma(\sigma_8)\sim 0.01$).

We thus include Gaussian priors 
$\sigma_p(\sigma_8) = \sigma_p(\Omega_m)=0.01$ in the analysis finding
a significant improvement
in the constraints on $M-c$ relation.
Uncertainties on $c_0$ reduce to $\sim10\%$ for both  
$snr=4$ and $snr=6$ while $\sigma_{\ln c}$ is reasonably (marginally) 
constrained for $snr=6 ~(4)$. Despite the large confidence interval,
constraints on $\alpha$ are however competitive with those
from strong lensing and X--ray observations 
(se e.g. \cite{johnston07,mandelbaum08,ettori11,sereno}). 

In fig. \ref{fm} we display the 2D $68\%$ and $95\%$ marginalized confidence
regions for different pairs of parameters for $snr^*=4$ and 
no priors added. Fig. \ref{fmprior} shows how the Fisher ellipses for 
$c_0$, $\alpha$ and $\sigma_{\ln c}$ are reduced 
after placing the priors $\sigma_p(\sigma_8)=\sigma_p(\Omega_m)=0.01$. 
Finally, the 2D constraints for $snr^*=4$ 
and $snr^*=6$ are compared in fig. \ref{fmprior2}.

\begin{figure}[]
\begin{center}
\includegraphics[angle=-90,scale=.30]{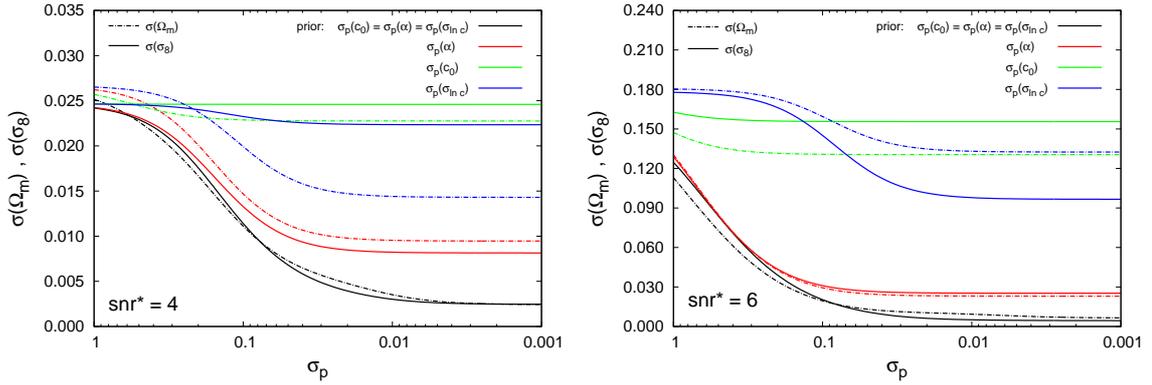}
\caption{Fisher matrix forecasted $1-\sigma$ uncertainties on $\sigma_8$ and $\Omega_m$ at the varying of the priors on $M-c$ relation parameters. Results 
are given for $snr^* =4$ ({\it left panel}) and $6$ ({\it right panel}).}
\label{sigcosm}
\end{center}
\end{figure}

On the other hand, it is also interesting to investigate the power of the
WL peak statistics to constrain the cosmological parameters when one or more 
$M-c$ relation parameters are known with good accuracy. 
To this aim, we consider the following cases where a prior $\sigma_p$ 
is added on:
i) $c_0$, ii) $\alpha$, iii) $\sigma_{\ln c}$, 
iv) all the $M-c$ relation parameters. 
Fig. \ref{sigcosm} shows the uncertainties on $\Omega_m$ and $\sigma_8$ at 
the varying of $\sigma_p$ for the cases $snr^*=4$ and $6$.
Results can be understood inspecting the correlation coefficients listed 
in table \ref{t2}. Among the $M-c$ relation parameters, the slope $\alpha$ 
is completely (anti)correlated with ($\sigma_8$) $\Omega_m$ so that 
it just suffices to know $\alpha$ with a sufficient precision to significantly 
reduce the cosmological parameter uncertainties. 
For our fiducial model and $snr^*=4$, constraints on $\sigma_8$ and 
$\Omega_m$ improve of about $\sim 50\%$($\sim 70\%$) 
(with respect to the values reported in the first row of table \ref{t3}) 
if $\alpha$ is known with an uncertainty of 
$\sigma(\alpha)/\alpha \simeq 1$($<0.1$). Similar values are obtained 
for $snr^*=6$. On the contrary, given their lower correlations,
even with a perfect knowledge of  
$c_0$ or $\sigma_{\ln c}$ limits on cosmological parameters are only 
modestly/scarcely reduced. 
The last two rows of table \ref{t3} show the uncertainties on 
$\Omega_m$ and $\sigma_8$ in the ideal case in which 
the $M-c$ relation is perfectly known. Although these numbers are 
too good and too optimistic and can not happen for real measurements, 
the above discussion highlights the importance of studies of the halo 
concentration parameters (in particular the slope $\alpha$) for 
future surveys.

\begin{figure}[]
\begin{center}
\includegraphics[angle=-90,scale=.30]{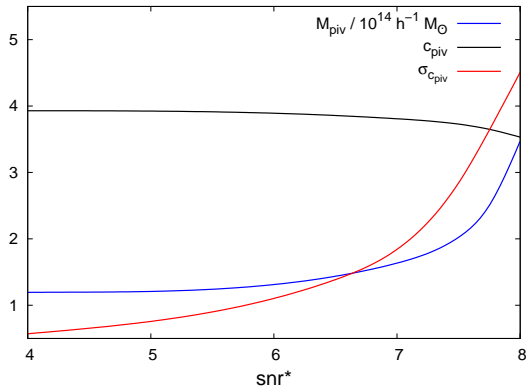}
\caption{$M_{\textrm {piv}}$, $c_{\textrm {piv}}$, and $\sigma_{c_{\textrm {piv}}}$ as a 
function of $snr^*$ (see text for details).}
\label{piv}
\end{center}
\end{figure}

Finally, it is reasonable to suppose that, the $M_{ap}$ statistics 
can probe the inner structure of halos only for sufficiently massive halos. 
It is then interesting to investigate which mass scale can better constrain 
the concentration parameter. To this aim we follow \cite{zentner} (see also
\cite{eis,hujain} for 
the same approach in a different contest) and define a pivot mass 
$M_{\textrm {piv}}$ for the $M-c$ relation. 
We then choose $M_{\textrm {piv}}$ such that the 
relative error: 
\begin{eqnarray} 
\frac{\sigma_{c_{\textrm{piv}}}^2}{c_{\textrm{piv}}^2} &=& \frac{1}{c_{\textrm{piv}}^2}
\sum_{i,j} \frac{\partial c_{\textrm{piv}}}{\partial p_i}C_{ij}
\frac{\partial c_{\textrm{piv}}}{\partial  p_j} \nonumber \\ 
%&=& \frac {[F^{-1}]_{c_0c_0}}{c_0^2} + 
%\left[ \ln \left( \frac{M_{\textrm{piv}}}{M_0} \right) \right]^2 [F^{-1}]_{\alpha \alpha} + 
%\frac{2}{c_0} \ln \left( \frac{M_{\textrm{piv}}}{M_0} \right) [F^{-1}]_{c_0 \alpha}\\
&=& \frac {C_{c_0c_0}}{c_0^2} + 
\left[ \ln \left( \frac{M_{\textrm{piv}}}{M_0} \right) \right]^2 C_{\alpha \alpha} + 
\frac{2}{c_0} \ln \left( \frac{M_{\textrm{piv}}}{M_0} \right) C_{c_0 \alpha}
\label{sigpiv}
\end{eqnarray}
(here $C_{ij}=(F^{-1})_{ij}$ is the parameter covariance matrix and $M_0=10^{14}h^{-1} M_\odot$, see (\ref{c200}))
on the rescaled normalization $c_{\textrm {piv}}=c_0 (M_{\textrm {piv}}/M_0)^\alpha$
is minimized (i.e. the errors on $c_{\textrm {piv}}$ and $\alpha$ are 
uncorrelated). This occurs when
$$
\ln \left( \frac {M_{\textrm {piv}}}{M_0} \right) =  -\frac{1}{c_0} \frac{C_{c_0 \alpha}}{C_{\alpha \alpha}} 
$$
yielding
$$
\frac{\sigma_{c_{\textrm{piv}}}^2}{c_{\textrm{piv}}^2} = \frac{1}{c_0^2} \left[
C_{c_0c_0} - \frac{C_{c_0 \alpha}^2}{C_{\alpha \alpha}} \right]
$$
The values of $M_{\textrm {piv}}$, $c_{\textrm {piv}}$ and 
$\sigma_{\textrm {piv}}$ as a function of $snr^*$ are displayed in 
fig. \ref{piv}. 
As the figure shows, $M_{\textrm {piv}}$ is quite insensitive
to the peak height threshold
up to  $snr^* \simeq 7$, 
ranging from $\sim 1.2$ to 
$\sim 1.5 \cdot 10^{14} h^{-1}M_\odot $.
Then, it rapidly increases up to 
$\sim 3.5 \cdot 10^{14} h^{-1}M_\odot $ at $snr^*=8$. This is probably due
to the fact that, for such high thresholds, the signal is mostly due to 
more massive halos which, unlike the less massive ones, feel
the impact of the concentration scatter (see section \ref{PDF}).
Results for $snr^*=4$ and $6$ are very close to those reported in 
table \ref{t3} (first two rows) and are summarized in 
table \ref{tpiv}.


It is worth mentioning, that peak number counts depend on the adopted filter
function so that the constraints here reported can change if  a
different filter is used. Investigating which filter is optimal
for peaks detection is outside the aims of this work. We only mention
that a filter which performs better than the one used here, 
would increase the number of detections improving 
the counts statistics and consequently the constraints reported in table \ref{t3}.

\section{Discussion and conclusions}
\label{concl}

Halo masses and concentrations have been studied extensively, by means of 
N--body simulations as well as observationally, during the last decade. 
Nevertheless, the exact form of the $M-c$ relation is still widely debated.
One of the most promising method to estimate masses and concentrations 
relies on gravitational lensing, in particular shear peak counts from 
near--future WL surveys seem a promising and complementary method to delineate
the features of the $M-c$ relation, as pointed out in this work.

The method for shear peak detection we have considered here relies on 
the $M_{ap}$ statistics. Hence, we have firstly investigated the impact 
of the $M-c$ relation on the PDF of $M_{ap}$.
In particular, we have provided a prescription to properly take into account
the $M-c$ relation dispersion, always disregarded in previous work, 
in the PDF. We have then shown that such scatter can 
affect significantly the PDF and, consequently, the $snr$ for peak detection.

Secondly, we have investigated the impact of the $M-c$ relation
on peaks number counts as compared to that of the cosmological 
parameters $\sigma_8$ and $\Omega_m$. 

Finally, we have performed a Fisher matrix analysis 
in order to assess the capability of an Euclid--like survey 
in constraining the $M-c$ relation and cosmological parameters by using
peak number counts.
We have found that: i) peak function alone
provides constraints on $\sigma_8$ 
and $\Omega_m$ which are competitive with those obtained from different probes.
On the other hand, only the normalization $c_0$ of the $M-c$ relation is 
reasonably (but not so strongly) constrained if peaks
with $snr>4$ are considered;
ii) adding prior informations on $\sigma_8$ and $\Omega_m$ as
inferred from current/near--future data, constraints on $M-c$ relation 
improve significantly so that peak statistics seems helpful in 
discriminating among the wide variety of $M-c$ relation fits found in 
literature although no substantial new information on the slope $\alpha$ is 
inferred.

Although results are quite encouraging, 
it is worth stressing that the Fisher matrix analysis 
here presented disregards the effects of the halo sample variance on the 
error estimates, only Poisson noise being considered.
Nevertheless, sample variance might be significant for low mass halos
at low redshifts where their abundance is much higher thus reducing 
the Poisson noise. To yield more accurate predictions,
the covariance matrix for number counts $C_{ij}$ 
\footnote{it should not be confused with the parameter covariance matrix 
of eq. (\ref{sigpiv}) for which we use the same simbol.}, entering the FM expression,
should then be written as: $C_{ij} = P_{ij}+S_{ij}$ where
$P_{ij}=\delta_{ij}N_i^{fid}$ and
$S_{ij}=N_i^{fid}N_j^{fid}b_ib_j \sigma_{ij}^{DM}$ are 
the Poissonian and sample variance contributions respectively 
(here $N_i^{fid}$ and $b_i$ are the expected number of peaks and the 
mean halo bias in the $i$--th redshift bin, while 
$\sigma_{ij}^{DM}$ is the covariance of the dark matter fluctuations 
between the bins $i$ and $j$, se e.g. \cite{hukra,takada,valageas} 
for more details).

Sample variance effects on halo number counts have been considered in
previous works, e.g. \cite{hukra,takada,valageas,lima,masamune}. 
In particular,  \cite{hukra,valageas} conclude that sample variance is 
generally comparable or greater than Poisson noise for number counts
above mass threshold $M_{th} \lesssim 1-4 \cdot 10^{14} h^{-1} M_\odot$
depending on the survey width and depth. 
In our case, the mass thresholds corresponding to given $snr^*$ and $z$ can be 
deduced from the middle panel of fig. \ref{snzM}. 
According to  \cite{hukra,valageas}, for $snr^* \geq 4$
we expect sample variance to have only moderate effects in the redshift range 
$\Delta z=0.15-0.45$ where the counts signal is maximum, becoming negligible
at the increasing of $z$. 
For a raw estimate, we can assume $C_{ij}$ to be diagonal (i.e.
halo correlation length smaller than the bin width) and 
$S_{ij}=P_{ij}$ ($S_{ij}=2P_{ij}$) independently
of the redshift. This yields a
worsening of the parameter constraints by a factor $\sqrt 2$ ($\sqrt 3$).

However,  \cite{hukra,takada,valageas} neglect the information 
arising from the cosmological dependence of the sample variance. 
When this information is included and considered as ``signal'',
it could cause a smaller degradation in the constraints or, in case,
even improvements \cite{lima,masamune}. 
In this case, the full FM can be well approximated
as:

$$
F_{ij} = {\bf N}^T_{,i} {\bf C}^{-1} {\bf N}_{,j} +
\frac{1}{2} {\textrm {Tr}} \left[{\bf C}^{-1}{\bf S}_{,i}{\bf C}^{-1}{\bf S}_{,j}
\right]+ \frac{\delta_{ij}}{\sigma_p^2(p_i)} 
$$
where
${\bf N}=(N^{fid}_1,...,N^{fid}_{n_{bin}})$ and $_{,i}$ denotes the derivative 
with respect to the parameter $p_i$. 
It is worth stressing that the second term, neglected in 
\cite{hukra,takada,valageas}, beside
the information about the cosmological dependence 
of the sample variance, also contains
information about its dependence on $M-c$ relation trough the
derivative of $N_i^{fid}$ and possibly of $b_i$.
Indeed, it was pointed out in \cite{wechsler,jingsu} that halo bias
could depend not only on the halo mass but also on additional halo properties, 
among which the concentration. 

Adding sample variance information 
could then, in principle, provide an opportunity
to improve measurements of halo concentration (and cosmological parameters) 
with WL peak statistics through self--calibration techniques 
for $M-c$ relation similar to those for mass--observable relation 
(see e.g. \cite{lima,masamune}). 
This intriguing point, which deserve a detailed analysis, is left
for future works.

A second issue concerns the assumption
that detected peaks can be properly assigned to redshift bins.
The redshift information, however, can not be extracted from WL data only,
but additional data need to be included. In order to split the peak sample 
in redshift bins, optical finders (see e.g. \cite{postman,koester,milke,bella})
could serve the purpose since they return reasonable 
redshift estimates, especially because they can be applied on the same 
optical data retrieved for lensing. Although the correlation between optical 
detections and WL peaks is not trivial, in a Euclid--like survey, 
one can look at the position of the peaks in the optical images, 
and then deduce the peak redshift $z$ from 
the redshifts of the galaxies closest to the peak position.
In a first approximation, we can suppose that this method provides
a Gaussian probability distribution function for $z$ 
with negligible bias and variance $\sigma_z = \sigma_0(1+z)$, and then
assume a peak to be correctly assigned to a redshift bin 
roughly asking that the $3-\sigma$ uncertainty 
on $z$ is smaller than the bin width. 

For our assumed value $\Delta z = 0.1$, this translates in 
$\sigma_0 \leq 0.03$, a precision which could be likely achieved 
if $z$ is spectroscopically measured ($\sigma_0 \simeq 0.001$ according to the
Euclid red book), but could be too demanding if one relies on 
photometric redshift methods ($\sigma_0 \simeq 0.05$ for
Euclid). In this second case, one should add a non Poissonian uncertainty
on $N_{halo}(snr > snr^*, z, \Delta z)$
in the above Fisher matrix analysis and resort to the pull statistics
\cite{campanelli}. 
On the other hand, it is likely that the precision of the inferred  
redshift also depends on the $snr$ of peak so that the net effect 
should be included in the analysis by convolving the 
theoretically computed $N_{halo}$ 
with an empirically determined selection function. 

A further issue concerns the assumption of NFW spherical halos.
Real halos are not spherical and not all relaxed so that deviations from the 
universal NFW profile are expected. The effects of the diversity of 
dark matter distributions in individual halos on peak counts has been 
investigated in \cite{gruen,hamana12} by means of numerical simulations and 
analytic methods. 
The noise originated from halo shape is found to be comparable
to the statistical noise discussed in section \ref{pdf}. Furthermore,
halo orientations cause a systematic bias in the peak heights.
In order to include these effects in the theoretical predictions, 
one should consider a triaxial halo model (see \cite{jingsuto})
which is shown to work quite well \cite{hamana12}.
  
It is also worth mentioning that, in the FM analysis, we 
have only investigated a subset of the full parameter space.
Here, we were mainly interested in how peak statistics constrain
the $M-c$ relation so that those parameters, as $H_0$ and $n_s$, which 
have a minor impact on peak number counts and are expected to be scarcely 
constrained, were held fixed to their 
fiducial values. Nevertheless, when dealing with real data, the full
parameter space should be considered. 
Allowing for a large number of parameters 
to be varied introduces further degeneracies widening the confidence 
regions. This degradation in the 
constraining power, can then be compensated by complementing 
peak statistics with further datasets, e.g. SNeIa and CMB.

Investigating the above issue is however outside our aims here.

\acknowledgments
RM is grateful to S. Bonometto for carefully reading the manuscript, helpful 
comments and suggestions. 
G. La Vacca is also thanked for useful discussions.
 
After this paper has been completed, we became aware of a very similar work by
V. Cardone, S. Camera, M. Sereno, G. Covone, R. Maoli and R. Scaramella 
\cite{cardonemc}. We thank them for sharing their draft paper and 
useful discussions.


\begin{thebibliography}{99}

\bibitem{Hett07}
M.~{Hetterscheidt}, P.~{Simon}, M.~{Schirmer}, H.~{Hildebrandt},
  T.~{Schrabback}, T.~{Erben}, and P.~{Schneider}, {\it {GaBoDS: The
  Garching-Bonn deep survey. VII. Cosmic shear analysis}},  {\em \aap} {\bf
  468} (2007) 859--876, [\href{http://xxx.lanl.gov/abs/astro-ph/0606571}{{\tt
  astro-ph/0606571}}].

\bibitem{Hoekstra06}
H.~{Hoekstra}, Y.~{Mellier}, L.~{van Waerbeke}, E.~{Semboloni}, L.~{Fu}, M.~J.
  {Hudson}, L.~C. {Parker}, I.~{Tereno}, and K.~{Benabed}, {\it {First Cosmic
  Shear Results from the Canada-France-Hawaii Telescope Wide Synoptic Legacy
  Survey}},  {\em \apj} {\bf 647} (2006) 116--127,
  [\href{http://xxx.lanl.gov/abs/astro-ph/0511089}{{\tt astro-ph/0511089}}].

\bibitem{Semboloni06}
E.~{Semboloni}, Y.~{Mellier}, L.~{van Waerbeke}, H.~{Hoekstra}, I.~{Tereno},
  K.~{Benabed}, S.~D.~J. {Gwyn}, L.~{Fu}, M.~J. {Hudson}, R.~{Maoli}, and L.~C.
  {Parker}, {\it {Cosmic shear analysis with CFHTLS deep data}},  {\em \aap}
  {\bf 452} (2006) 51--61,
  [\href{http://xxx.lanl.gov/abs/astro-ph/0511090}{{\tt astro-ph/0511090}}].

\bibitem{euclid1}

R.~{Laurejis} et al., 
{\it Euclid Definition Study Report}, (2011),
[\href{http://xxx.lanl.gov/abs/1110.3193 }{{\tt arXiv:1110.3193}}].

\bibitem{euclid2}
L.~{Amendola} et al.,
{\it Cosmology and Fundamental Physics with the Euclid Satellite},
Living Reviews in Relativity {\bf 16} (2013),
[\href{http://xxx.lanl.gov/abs/1206.1225}{{\tt arXiv:1206.1225}}].

\bibitem{kruse99}
G.~{Kruse} and P.~{Schneider}, {\it {Statistics of dark matter haloes expected
  from weak lensing surveys}},  {\em \mnras} {\bf 302} (1999) 821--829,
  [\href{http://xxx.lanl.gov/abs/astro-ph/9806071}{{\tt astro-ph/9806071}}].

\bibitem{Reblinsky99}
K.~{Reblinsky}, G.~{Kruse}, B.~{Jain}, and P.~{Schneider}, {\it {Cosmic shear
  and halo abundances: analytical versus numerical results}},  {\em \aap} {\bf
  351} (1999) 815--826, [\href{http://xxx.lanl.gov/abs/astro-ph/9907250}{{\tt
  astro-ph/9907250}}].

\bibitem{Dietrich10}
J.~P. {Dietrich} and J.~{Hartlap}, {\it {Cosmology with the shear-peak
  statistics}},  {\em \mnras} {\bf 402} (2010) 1049--1058,
  [\href{http://xxx.lanl.gov/abs/0906.3512}{{\tt arXiv:0906.3512}}].

\bibitem{Hoekstra01}
H.~{Hoekstra}, {\it {The effect of distant large scale structure on weak
  lensing mass estimates}},  {\em \aap} {\bf 370} (2001) 743--753,
  [\href{http://xxx.lanl.gov/abs/astro-ph/0102368}{{\tt astro-ph/0102368}}].

\bibitem{Hoekstra03}
H.~{Hoekstra}, {\it {How well can we determine cluster mass profiles from weak
  lensing?}},  {\em \mnras} {\bf 339} (2003) 1155--1162,
  [\href{http://xxx.lanl.gov/abs/astro-ph/0208351}{{\tt astro-ph/0208351}}].

\bibitem{Hoekstra11}
H.~{Hoekstra}, J.~{Hartlap}, S.~{Hilbert}, and E.~{van Uitert}, {\it {Effects
  of distant large-scale structure on the precision of weak lensing mass
  measurements}},  {\em \mnras} {\bf 412} (2011) 2095--2103,
  [\href{http://xxx.lanl.gov/abs/1011.1084}{{\tt arXiv:1011.1084}}].

\bibitem{hamana04}
T.~{Hamana}, M.~{Takada}, and N.~{Yoshida}, {\it {Searching for massive
  clusters in weak lensing surveys}},  {\em \mnras} {\bf 350} (May, 2004)
  893--913, [\href{http://xxx.lanl.gov/abs/astro-ph/0310607}{{\tt
  astro-ph/0310607}}].

\bibitem{marian06}
L.~{Marian} and G.~M. {Bernstein}, {\it {Dark energy constraints from
  lensing-detected galaxy clusters}},  {\em \prd} {\bf 73} (2006), no.~12
  123525, [\href{http://xxx.lanl.gov/abs/astro-ph/0605746}{{\tt
  astro-ph/0605746}}].

\bibitem{maturi10}
M.~{Maturi}, C.~{Angrick}, F.~{Pace}, and M.~{Bartelmann}, {\it {An analytic
  approach to number counts of weak-lensing peak detections}},  {\em \aap} {\bf
  519} (2010) A23, [\href{http://xxx.lanl.gov/abs/0907.1849}{{\tt
  arXiv:0907.1849}}].

\bibitem{maturi11}
M.~{Maturi}, C.~{Fedeli}, L.~{Moscardini},
{\it {Imprints of primordial non-Gaussianity on the number counts of cosmic 
shear peaks}},
{\em \mnras} {\bf 416} (2011) 2527-2538,
[\href{http://xxx.lanl.gov/abs/1101.4175}{{\tt arXiv:1101.4175}}].

\bibitem{cardone}
V.~F. {Cardone}, S.~{Camera}, R.~{Mainini}, A.~{Romano}, A.~{Diaferio}, 
R.~{Maoli}, R.~{Scaramella},
{\it Weak lensing peak count as a probe of f(R) theories},
{\em \mnras} {\bf 430} (2013) 2896-2909, 
[\href{http://xxx.lanl.gov/abs/1204.3148}{{\tt arXiv:1204.3148}}].

\bibitem{hett05}
M.~{Hetterscheidt}, T.~{Erben}, P.~{Schneider}, R.~{Maoli}, L.~{van Waerbeke},
  and Y.~{Mellier}, {\it {Searching for galaxy clusters using the aperture mass
  statistics in 50 VLT fields}},  {\em \aap} {\bf 442} (2005) 43--61,
  [\href{http://xxx.lanl.gov/abs/astro-ph/0504635}{{\tt astro-ph/0504635}}].

\bibitem{dahle06}
H.~{Dahle}, {\it {The Cluster Mass Function from Weak Gravitational Lensing}},
  {\em \apj} {\bf 653} (2006) 954--962,
  [\href{http://xxx.lanl.gov/abs/astro-ph/0608480}{{\tt astro-ph/0608480}}].

\bibitem{wittman06}
D.~{Wittman}, I.~P. {Dell'Antonio}, J.~P. {Hughes}, V.~E. {Margoniner}, J.~A.
  {Tyson}, J.~G. {Cohen}, and D.~{Norman}, {\it {First Results on
  Shear-selected Clusters from the Deep Lens Survey: Optical Imaging,
  Spectroscopy, and X-Ray Follow-up}},  {\em \apj} {\bf 643} (2006) 128--143,
  [\href{http://xxx.lanl.gov/abs/astro-ph/0507606}{{\tt astro-ph/0507606}}].

\bibitem{gavazzi07}
R.~{Gavazzi} and G.~{Soucail}, {\it {Weak lensing survey of galaxy clusters in
  the CFHTLS Deep}},  {\em \aap} {\bf 462} (2007) 459--471,
  [\href{http://xxx.lanl.gov/abs/astro-ph/0605591}{{\tt astro-ph/0605591}}].

\bibitem{schirmer07}
M.~{Schirmer}, T.~{Erben}, M.~{Hetterscheidt}, and P.~{Schneider}, {\it
  {GaBoDS: the Garching-Bonn Deep Survey. IX. A sample of 158 shear-selected
  mass concentration candidates}},  {\em \aap} {\bf 462} (2007) 875--887,
  [\href{http://xxx.lanl.gov/abs/astro-ph/0607022}{{\tt astro-ph/0607022}}].

\bibitem{miyazaki07}
S.~{Miyazaki}, T.~{Hamana}, R.~S. {Ellis}, N.~{Kashikawa}, R.~J. {Massey},
  J.~{Taylor}, and A.~{Refregier}, {\it {A Subaru Weak-Lensing Survey. I.
  Cluster Candidates and Spectroscopic Verification}},  {\em \apj} {\bf 669}
  (2007) 714--728, [\href{http://xxx.lanl.gov/abs/0707.2249}{{\tt
  arXiv:0707.2249}}].

\bibitem{berge08}
J.~{Berg{\'e}}, F.~{Pacaud}, A.~{R{\'e}fr{\'e}gier}, R.~{Massey}, M.~{Pierre},
  A.~{Amara}, M.~{Birkinshaw}, S.~{Paulin-Henriksson}, G.~P. {Smith}, and
  J.~{Willis}, {\it {Combined analysis of weak lensing and X-ray blind
  surveys}},  {\em \mnras} {\bf 385} (2008) 695--707,
  [\href{http://xxx.lanl.gov/abs/0712.3293}{{\tt arXiv:0712.3293}}].

\bibitem{abate09}
A.~{Abate}, D.~{Wittman}, V.~E. {Margoniner}, S.~L. {Bridle}, P.~{Gee}, J.~A.
  {Tyson}, and I.~P. {Dell'Antonio}, {\it {Shear-selected Clusters from the
  Deep Lens Survey. III. Masses from Weak Lensing}},  {\em \apj} {\bf 702}
  (2009) 603--613, [\href{http://xxx.lanl.gov/abs/0904.2185}{{\tt
  arXiv:0904.2185}}].

\bibitem{shan11}
H.~{Shan}, J.-P. {Kneib}, C.~{Tao}, Z.~{Fan}, M.~{Jauzac}, M.~{Limousin},
  R.~{Massey}, J.~{Rhodes}, K.~{Thanjavur}, and H.~J. {McCracken}, {\it {Weak
  lensing measurement of galaxy clusters in the CFHTLS-Wide survey}}, 
  (2011) [\href{http://xxx.lanl.gov/abs/1108.1981}{{\tt
  arXiv:1108.1981}}].

\bibitem{okabe10}
N.~{Okabe}, Y.-Y. {Zhang}, A.~{Finoguenov}, M.~{Takada}, G.~P. {Smith},
  K.~{Umetsu}, and T.~{Futamase}, {\it {LoCuSS: Calibrating Mass-observable
  Scaling Relations for Cluster Cosmology with Subaru Weak-lensing
  Observations}},  {\em \apj} {\bf 721} (2010) 875--885,
  [\href{http://xxx.lanl.gov/abs/1007.3816}{{\tt arXiv:1007.3816}}].

\bibitem{israel10}
H.~{Israel}, T.~{Erben}, T.~H. {Reiprich}, A.~{Vikhlinin}, H.~{Hildebrandt},
  D.~S. {Hudson}, B.~A. {McLeod}, C.~L. {Sarazin}, P.~{Schneider}, and Y.-Y.
  {Zhang}, {\it {The 400d Galaxy Cluster Survey weak lensing programme. I.
  MMT/Megacam analysis of CL0030+2618 at z = 0.50}},  {\em \aap} {\bf 520}
  (2010) A58, [\href{http://xxx.lanl.gov/abs/0911.3111}{{\tt
  arXiv:0911.3111}}].

\bibitem{S96}
P.~{Schneider}, {\it {Detection of (dark) matter concentrations via weak
  gravitational lensing}},  {\em \mnras} {\bf 283} (1996) 837--853,
  [\href{http://xxx.lanl.gov/abs/astro-ph/9601039}{{\tt astro-ph/9601039}}].

\bibitem{king11}
L.~J. {King} and J.~M.~G. {Mead}, {\it {The mass-concentration relationship of
  virialized haloes and its impact on cosmological observables}},  {\em \mnras}
  {\bf 416} (2011) 2539--2549, [\href{http://xxx.lanl.gov/abs/1105.3155}{{\tt
  arXiv:1105.3155}}].

\bibitem{B01}
J.~S. {Bullock}, T.~S. {Kolatt}, Y.~{Sigad}, R.~S. {Somerville}, A.~V.
  {Kravtsov}, A.~A. {Klypin}, J.~R. {Primack}, and A.~{Dekel}, {\it {Profiles
  of dark haloes: evolution, scatter and environment}},  {\em \mnras} {\bf 321}
  (2001) 559--575, [\href{http://xxx.lanl.gov/abs/astro-ph/9908159}{{\tt
  astro-ph/9908159}}].

\bibitem{eke01}
V.~R. {Eke}, J.~F. {Navarro}, and M.~{Steinmetz}, {\it {The Power Spectrum
  Dependence of Dark Matter Halo Concentrations}},  {\em \apj} {\bf 554} (2001)
  114--125, [\href{http://xxx.lanl.gov/abs/astro-ph/0012337}{{\tt
  astro-ph/0012337}}].

\bibitem{Comerford07}
J.~M. {Comerford} and P.~{Natarajan}, {\it {The observed concentration-mass
  relation for galaxy clusters}},  {\em \mnras} {\bf 379} (July, 2007)
  190--200, [\href{http://xxx.lanl.gov/abs/astro-ph/0703126}{{\tt
  astro-ph/0703126}}].

\bibitem{neto07}
A.~F. {Neto}, L.~{Gao}, P.~{Bett}, S.~{Cole}, J.~F. {Navarro}, C.~S. {Frenk},
  S.~D.~M. {White}, V.~{Springel}, and A.~{Jenkins}, {\it {The statistics of
  {$\Lambda$} CDM halo concentrations}},  {\em \mnras} {\bf 381} (2007)
  1450--1462, [\href{http://xxx.lanl.gov/abs/0706.2919}{{\tt
  arXiv:0706.2919}}].

\bibitem{gao08}
L.~{Gao}, J.~F. {Navarro}, S.~{Cole}, C.~S. {Frenk}, S.~D.~M. {White},
  V.~{Springel}, A.~{Jenkins}, and A.~F. {Neto}, {\it {The redshift dependence
  of the structure of massive {$\Lambda$} cold dark matter haloes}},  {\em
  \mnras} {\bf 387} (2008) 536--544,
  [\href{http://xxx.lanl.gov/abs/0711.0746}{{\tt arXiv:0711.0746}}].

\bibitem{duffy08}
A.~R. {Duffy}, J.~{Schaye}, S.~T. {Kay}, and C.~{Dalla Vecchia}, {\it {Dark
  matter halo concentrations in the Wilkinson Microwave Anisotropy Probe year 5
  cosmology}},  {\em \mnras} {\bf 390} (2008) L64--L68,
  [\href{http://xxx.lanl.gov/abs/0804.2486}{{\tt arXiv:0804.2486}}].

\bibitem{maccio08}
A.~V. {Macci{\`o}}, A.~A. {Dutton}, and F.~C. {van den Bosch}, {\it
  {Concentration, spin and shape of dark matter haloes as a function of the
  cosmological model: WMAP1, WMAP3 and WMAP5 results}},  {\em \mnras} {\bf 391}
  (2008) 1940--1954, [\href{http://xxx.lanl.gov/abs/0805.1926}{{\tt
  arXiv:0805.1926}}].

\bibitem{mandelbaum08}
R.~{Mandelbaum}, U.~{Seljak}, and C.~M. {Hirata}, {\it {A halo
  mass--concentration relation from weak lensing}},  {\em \jcap} {\bf 8}
  (2008) 6, [\href{http://xxx.lanl.gov/abs/0805.2552}{{\tt arXiv:0805.2552}}].

\bibitem{oguri09}
M.~{Oguri}, J.~F. {Hennawi}, M.~D. {Gladders}, H.~{Dahle}, P.~{Natarajan},
  N.~{Dalal}, B.~P. {Koester}, K.~{Sharon}, and M.~{Bayliss}, {\it {Subaru Weak
  Lensing Measurements of Four Strong Lensing Clusters: Are Lensing Clusters
  Overconcentrated?}},  {\em \apj} {\bf 699} (2009) 1038--1052,
  [\href{http://xxx.lanl.gov/abs/0901.4372}{{\tt arXiv:0901.4372}}].

\bibitem{klypin11}
A.~A. {Klypin}, S.~{Trujillo-Gomez}, and J.~{Primack}, {\it {Dark Matter Halos
  in the Standard Cosmological Model: Results from the Bolshoi Simulation}},
  {\em \apj} {\bf 740} (2011) 102,
  [\href{http://xxx.lanl.gov/abs/1002.3660}{{\tt arXiv:1002.3660}}].

\bibitem{prada11}
F.~{Prada}, A.~A. {Klypin}, A.~J. {Cuesta}, J.~E. {Betancort-Rijo}, and
  J.~{Primack}, {\it {Halo concentrations in the standard LCDM cosmology}},
  (2011) [\href{http://xxx.lanl.gov/abs/1104.5130}{{\tt
  arXiv:1104.5130}}].

\bibitem{meneghetti13}
M.~{Meneghetti} and E.~{Rasia},
{\it {Reconciling extremely different concentration-mass relations}},
submitted to {\em \mnras}  (2013), 
[\href{http://xxx.lanl.gov/abs/1303.6158}{{\tt arXiv:1303.6158}}].

\bibitem{NFW}
J.~F. {Navarro}, C.~S. {Frenk}, and S.~D.~M. {White}, {\it {The Structure of
  Cold Dark Matter Halos}},  {\em \apj} {\bf 462} (1996) 563,
  [\href{http://xxx.lanl.gov/abs/astro-ph/9508025}{{\tt astro-ph/9508025}}].

\bibitem{dolag04}
K.~{Dolag}, M.~{Bartelmann}, F.~{Perrotta}, C.~{Baccigalupi}, L.~{Moscardini},
  M.~{Meneghetti}, and G.~{Tormen}, {\it {Numerical study of halo
  concentrations in dark-energy cosmologies}},  {\em \aap} {\bf 416} (2004)
  853--864, [\href{http://xxx.lanl.gov/abs/astro-ph/0309771}{{\tt
  astro-ph/0309771}}].

\bibitem{zhao09}
D.~H. {Zhao}, Y.~P. {Jing}, H.~J. {Mo}, and G.~{B{\"o}rner}, {\it {Accurate
  Universal Models for the Mass Accretion Histories and Concentrations of Dark
  Matter Halos}},  {\em \apj} {\bf 707} (2009) 354--369,
  [\href{http://xxx.lanl.gov/abs/0811.0828}{{\tt arXiv:0811.0828}}].

\bibitem{Jing00}
Y.~P. {Jing}, {\it {The Density Profile of Equilibrium and Nonequilibrium Dark
  Matter Halos}},  {\em \apj} {\bf 535} (2000) 30--36,
  [\href{http://xxx.lanl.gov/abs/astro-ph/9901340}{{\tt astro-ph/9901340}}].

\bibitem{MC11}
J.~C. {Mu{\~n}oz-Cuartas}, A.~V. {Macci{\`o}}, S.~{Gottl{\"o}ber}, and A.~A.
  {Dutton}, {\it {The redshift evolution of {$\Lambda$} cold dark matter halo
  parameters: concentration, spin and shape}},  {\em \mnras} {\bf 411} (2011)
  584--594, [\href{http://xxx.lanl.gov/abs/1007.0438}{{\tt arXiv:1007.0438}}].

\bibitem{rines06}
K.~{Rines} and A.~{Diaferio}, {\it {CIRS: Cluster Infall Regions in the Sloan
  Digital Sky Survey. I. Infall Patterns and Mass Profiles}},  {\em \aj} {\bf
  132} (2006) 1275--1297, [\href{http://xxx.lanl.gov/abs/astro-ph/0602032}{{\tt
  astro-ph/0602032}}].

\bibitem{WL10}
R.~{Wojtak} and E.~L. {{\L}okas}, {\it {Mass profiles and galaxy orbits in
  nearby galaxy clusters from the analysis of the projected phase space}},
  {\em \mnras} {\bf 408} (2010) 2442--2456,
  [\href{http://xxx.lanl.gov/abs/1004.3771}{{\tt arXiv:1004.3771}}].

\bibitem{buote07}
D.~A. {Buote}, F.~{Gastaldello}, P.~J. {Humphrey}, L.~{Zappacosta}, J.~S.
  {Bullock}, F.~{Brighenti}, and W.~G. {Mathews}, {\it {The X-Ray
  Concentration-Virial Mass Relation}},  {\em \apj} {\bf 664} (2007) 123--134,
  [\href{http://xxx.lanl.gov/abs/astro-ph/0610135}{{\tt astro-ph/0610135}}].

\bibitem{SA07}
R.~W. {Schmidt} and S.~W. {Allen}, {\it {The dark matter haloes of massive,
  relaxed galaxy clusters observed with Chandra}},  {\em \mnras} {\bf 379}
  (2007) 209--221, [\href{http://xxx.lanl.gov/abs/astro-ph/0610038}{{\tt
  astro-ph/0610038}}].

\bibitem{johnston07}
D.~E. {Johnston}, E.~S. {Sheldon}, A.~{Tasitsiomi}, J.~A. {Frieman}, R.~H.
  {Wechsler}, and T.~A. {McKay}, {\it {Cross-Correlation Lensing: Determining
  Galaxy and Cluster Mass Profiles from Statistical Weak-Lensing
  Measurements}},  {\em \apj} {\bf 656} (2007) 27--41,
  [\href{http://xxx.lanl.gov/abs/astro-ph/0507467}{{\tt astro-ph/0507467}}].

\bibitem{broadhurst08}
T.~{Broadhurst}, K.~{Umetsu}, E.~{Medezinski}, M.~{Oguri}, and Y.~{Rephaeli},
  {\it {Comparison of Cluster Lensing Profiles with {$\Lambda$}CDM
  Predictions}},  {\em \apjl} {\bf 685} (2008) L9--L12,
  [\href{http://xxx.lanl.gov/abs/0805.2617}{{\tt arXiv:0805.2617}}].

\bibitem{vikhlinin09}
A.~{Vikhlinin}, A.~V. {Kravtsov}, R.~A. {Burenin}, H.~{Ebeling}, W.~R.
  {Forman}, A.~{Hornstrup}, C.~{Jones}, S.~S. {Murray}, D.~{Nagai},
  H.~{Quintana}, and A.~{Voevodkin}, {\it {Chandra Cluster Cosmology Project
  III: Cosmological Parameter Constraints}},  {\em \apj} {\bf 692} (2009)
  1060--1074, [\href{http://xxx.lanl.gov/abs/0812.2720}{{\tt
  arXiv:0812.2720}}].

\bibitem{ettori11}
S.~{Ettori}, F.~{Gastaldello}, A.~{Leccardi}, S.~{Molendi}, M.~{Rossetti},
  D.~{Buote}, and M.~{Meneghetti}, {\it {Mass profiles and c - M$_{DM}$
  relation in X-ray luminous galaxy clusters (Corrigendum)}},  {\em \aap} {\bf
  526} (2011) C1, [\href{http://xxx.lanl.gov/abs/1009.3266}{{\tt
  arXiv:1009.3266}}].

\bibitem{oguri05}
M.~{Oguri}, M.~{Takada}, K.~{Umetsu}, and T.~{Broadhurst}, {\it {Can the Steep
  Mass Profile of A1689 Be Explained by a Triaxial Dark Halo?}},  {\em \apj}
  {\bf 632} (2005) 841--846,
  [\href{http://xxx.lanl.gov/abs/astro-ph/0505452}{{\tt astro-ph/0505452}}].

\bibitem{sereno11}
M.~{Sereno} and K.~{Umetsu}, {\it {Weak- and strong-lensing analyses of the
  triaxial matter distribution of Abell 1689}},  {\em \mnras} {\bf 416} (2011)
  3187--3200, [\href{http://xxx.lanl.gov/abs/1105.4994}{{\tt
  arXiv:1105.4994}}].

\bibitem{giocoli}
C.~{Giocoli}, M.~{Meneghetti}, S.~{Ettori}, L.~{Moscardini},
{\it {Cosmology in 2D: the concentration-mass relation for galaxy clusters}},
{\em \mnras} {\bf 426} (2011) 1558--1573,
  [\href{http://xxx.lanl.gov/abs/1205.2375}{{\tt
  arXiv:1205.2375}}].

\bibitem{corless09}
V.~L. {Corless} and L.~J. {King}, {\it {Cosmology with the cluster mass
  function: mass estimators and shape systematics in large weak lensing
  surveys}},  {\em \mnras} {\bf 396} (2009) 315--324,
  [\href{http://xxx.lanl.gov/abs/0901.3434}{{\tt arXiv:0901.3434}}].

\bibitem{KC07}
L.~{King} and V.~{Corless}, {\it {Complex structures in galaxy cluster fields:
  implications for gravitational lensing mass models}},  {\em \mnras} {\bf 374}
  (2007) L37--L41, [\href{http://xxx.lanl.gov/abs/astro-ph/0610493}{{\tt
  astro-ph/0610493}}].

\bibitem{gnedin11}
O.~Y. {Gnedin}, D.~{Ceverino}, N.~Y. {Gnedin}, A.~A. {Klypin}, A.~V.
  {Kravtsov}, R.~{Levine}, D.~{Nagai}, and G.~{Yepes}, {\it {Halo Contraction
  Effect in Hydrodynamic Simulations of Galaxy Formation}},  
  (2011) [\href{http://xxx.lanl.gov/abs/1108.5736}{{\tt
  arXiv:1108.5736}}].

\bibitem{sereno10}
M.~{Sereno}, M.~{Lubini}, and P.~{Jetzer}, {\it {A multiwavelength strong
  lensing analysis of baryons and dark matter in the dynamically active cluster
  AC 114}},  {\em \aap} {\bf 518} (2010) A55,
  [\href{http://xxx.lanl.gov/abs/0904.0018}{{\tt arXiv:0904.0018}}].

\bibitem{fedeli11}
C.~{Fedeli}, {\it {The effects of baryonic cooling on the concentration-mass
  relation}}, (2011)
  [\href{http://xxx.lanl.gov/abs/1111.5780}{{\tt arXiv:1111.5780}}].

\bibitem{evrard96}
A.~E. {Evrard}, C.~A. {Metzler}, and J.~F. {Navarro}, {\it {Mass Estimates of
  X-Ray Clusters}},  {\em \apj} {\bf 469} (1996) 494,
  [\href{http://xxx.lanl.gov/abs/astro-ph/9510058}{{\tt astro-ph/9510058}}].

\bibitem{dolag05}
K.~{Dolag}, F.~{Vazza}, G.~{Brunetti}, and G.~{Tormen}, {\it {Turbulent gas
  motions in galaxy cluster simulations: the role of smoothed particle
  hydrodynamics viscosity}},  {\em \mnras} {\bf 364} (2005) 753--772,
  [\href{http://xxx.lanl.gov/abs/astro-ph/0507480}{{\tt astro-ph/0507480}}].

\bibitem{rasia06}
E.~{Rasia}, S.~{Ettori}, L.~{Moscardini}, P.~{Mazzotta}, S.~{Borgani},
  K.~{Dolag}, G.~{Tormen}, L.~M. {Cheng}, and A.~{Diaferio}, {\it {Systematics
  in the X-ray cluster mass estimators}},  {\em \mnras} {\bf 369} (2006)
  2013--2024, [\href{http://xxx.lanl.gov/abs/astro-ph/0602434}{{\tt
  astro-ph/0602434}}].

\bibitem{nagai07}
D.~{Nagai}, A.~{Vikhlinin}, and A.~V. ~{Kravtsov}, {\it {Testing X-Ray
  Measurements of Galaxy Clusters with Cosmological Simulations}},  {\em \apj}
  {\bf 655} (2007) 98--108,
  [\href{http://xxx.lanl.gov/abs/astro-ph/0609247}{{\tt astro-ph/0609247}}].

\bibitem{rasia13}
E.~{Rasia}, S.~{Borgani}, S.~{Ettori}, P.~{Mazzotta}, M.~{Meneghetti}
{\it {On the Discrepancy between Theoretical and X-Ray Concentration-Mass Relations for Galaxy Clusters}},
{\em \apj} in press (2013),
  [\href{http://xxx.lanl.gov/abs/1301.7476}{{\tt arXiv:1301.7476}}].

\bibitem{io1}
A. ~{Klypin}, A.~V. ~{Maccio'}, R. ~{Mainini}, S.~A. ~{Bonometto},
{\it Halo properties in models with dynamical Dark Energy}
{\em \apj} {\bf 599} (2003) 31--37,  
[\href{http://xxx.lanl.gov/abs/astro-ph/0303304}{{\tt astro-ph/0303304}}].

\bibitem{io2}
A.~V. ~{Maccio'}, C. ~{Quercellini}, R. ~{Mainini}, L. ~{Amendola}, 
S. ~A. ~{Bonometto}, {\it Coupled dark energy: Parameter constraints from N-body simulations}, {\em \prd} {\bf 69} (2004) 123516, 
[\href{http://xxx.lanl.gov/abs/astro-ph/0309671}{{\tt astro-ph/0309671}}].

\bibitem{baldi}
M. ~{Baldi}, V. ~{Pettorino}, G. ~{Robbers}, V.~{Springel},
{\it Hydrodynamical N-body simulations of coupled dark energy cosmologies},
{\em \mnras} {\bf 403} (2010) 1684B, 
[\href{http://xxx.lanl.gov/abs/0812.3901}{{\tt arXiv:0812.3901}}].

\bibitem{baldi2}
M. ~{Baldi}
{\it Time dependent couplings in the dark sector: from background evolution to nonlinear structure formation}
{\em \mnras} {\bf 411} (2011) 1077, 
[\href{http://xxx.lanl.gov/abs/1005.2188}{{\tt arXiv:1005.2188}}].

\bibitem{boni}
C.~{De Boni}, S.~{Ettori}, k.~{ Dolag}, L.~{Moscardini},
{\it Hydrodynamical simulations of galaxy clusters in dark energy cosmologies - II. c-M relation},
{\em \mnras} {\bf 428} (2013) 2921-2938,
[\href{http://xxx.lanl.gov/abs/1205.3163}{{\tt arXiv:1205.3163}}].

\bibitem{schirmer}
M.~{Schirmer}, T.~{Erben}, P.~{Schneider}, C.~{Wolf}, and K.~{Meisenheimer},
  {\it {GaBoDS: The Garching-Bonn Deep Survey. II. Confirmation of EIS cluster
  candidates by weak gravitational lensing}},  {\em \aap} {\bf 420} (2004)
  75--78, [\href{http://xxx.lanl.gov/abs/astro-ph/0401203}{{\tt
  astro-ph/0401203}}].

\bibitem{maturi05}
M.~{Maturi}, M.~{Meneghetti}, M.~{Bartelmann}, K.~{Dolag}, and L.~{Moscardini},
  {\it {An optimal filter for the detection of galaxy clusters through weak
  lensing}},  {\em \aap} {\bf 442} (2005) 851--860,
  [\href{http://xxx.lanl.gov/abs/astro-ph/0412604}{{\tt astro-ph/0412604}}].

\bibitem{HS05}
J.~F. {Hennawi} and D.~N. {Spergel}, {\it {Shear-selected Cluster Cosmology. 1:
  Tomography and Optimal Filtering}},  {\em \apj} {\bf 624} (2005) 59--79,
  [\href{http://xxx.lanl.gov/abs/astro-ph/0404349}{{\tt astro-ph/0404349}}].

\bibitem{KS93}
N.~{Kaiser} and G.~{Squires}, {\it {Mapping the dark matter with weak
  gravitational lensing}},  {\em \apj} {\bf 404} (1993) 441--450.

\bibitem{KSB95}
N.~{Kaiser}, G.~{Squires}, and T.~{Broadhurst}, {\it {A Method for Weak Lensing
  Observations}},  {\em \apj} {\bf 449} (1995) 460,
  [\href{http://xxx.lanl.gov/abs/astro-ph/9411005}{{\tt astro-ph/9411005}}].

\bibitem{JVW00}
B.~{Jain} and L.~{Van Waerbeke}, {\it {Statistics of Dark Matter Halos from
  Gravitational Lensing}},  {\em \apjl} {\bf 530} (2000) L1--L4,
  [\href{http://xxx.lanl.gov/abs/astro-ph/9910459}{{\tt astro-ph/9910459}}].

\bibitem{SWJK}
P.~{Schneider}, L.~{van Waerbeke}, B.~{Jain}, and G.~{Kruse}, {\it {A new
  measure for cosmic shear}},  {\em \mnras} {\bf 296} (1998) 873--892,
  [\href{http://xxx.lanl.gov/abs/astro-ph/9708143}{{\tt astro-ph/9708143}}].

\bibitem{gruen}
D. ~{Gruen}, G. ~M. ~{Bernstein}, T. ~Y. ~{Lam}, S. ~{Seitz},
{\it Optimizing weak lensing mass estimates for cluster profile uncertainty},
{\em \mnras} {\bf 416} (2011) 1392
 [\href{http://xxx.lanl.gov/abs/1104.2596}{{\tt arXiv:1104.2596}}].

\bibitem{marian11}
L.~{Marian}, S.~{Hilbert}, R.~E. {Smith}, P.~{Schneider}, and V.~{Desjacques},
  {\it {Measuring Primordial Non-gaussianity Through Weak-lensing Peak
  Counts}},  {\em \apjl} {\bf 728} (2011) L13,
  [\href{http://xxx.lanl.gov/abs/1010.5242}{{\tt arXiv:1010.5242}}].

\bibitem{hamana12}
T.~{Hamana}, M.~{Oguri}, M.~{Shirasaki}, M.~{Sato},
{\it Scatter and bias in weak lensing selected clusters},
{\em \mnras} {\bf 425} (2012) 2287--2298,
[\href{http://xxx.lanl.gov/abs/1204.6117}{{\tt arXiv:1204.6117}}].

\bibitem{fan}
Z.~{Fan}, H.~{Shan}, J.~{Liu}
{\it Noisy weak-lensing convergence peak statistics near clusters of galaxies and beyond},
{\em \apj} {\bf 719} (2010) 1408--1420
[\href{http://xxx.lanl.gov/abs/1006.5121}{{\tt arXiv:1006.5121}}].

\bibitem{bahe}
Y.~M. {Bahe'}, I.~G. {McCarthy}, L.~J. {King},
{\it Mock weak lensing analysis of simulated galaxy clusters: 
bias and scatter in mass and concentration},
{\em \mnras} {\bf 421} (2012) 1073--1088
[\href{http://xxx.lanl.gov/abs/1106.2046}{{\tt arXiv:1106.2046}}].

\bibitem{jingsuto}
Y.~P. {Jing}, Y.~{Suto},
{\it Triaxial Modeling of Halo Density Profiles with High-Resolution N-Body Simulations},
{\em \apj } {\bf 574} (2002 ) 538--553,
[\href{http://xxx.lanl.gov/abs/astro-ph/0202064}{{\tt astro-ph/0202064}}].

\bibitem{wmap9}
G.~{Hinshaw}  D.~{Larson}, E.~{Komatsu}, D.~N. {Spergel}, C.~L. {Bennett}, J.~{Dunkley}, M.~R. {Nolta}, M.~{Halpern}, R.~S. {Hill}, N.~{Odegard}, L.~{Page}, K.~M. {Smith}, J.~L. {Weiland}, B.~{Gold}, N.~{Jarosik}, A.~{Kogut}, M.~{Limon}, S.~S. {Meyer}, G.~S. {Tucker}, E.~{Wollack}, E.~L. {Wright},
{\it Nine-Year Wilkinson Microwave Anisotropy Probe (WMAP) Observations: Cosmological Parameter Results}, (2012),
[\href{http://xxx.lanl.gov/abs/1212.5226}{{\tt arXiv:1212.5226}}].

\bibitem{marian09}
L.~{Marian}, R.~E. {Smith}, G.~M. {Bernstein}
{\it The impact of correlated projections on weak lensing cluster counts}
{\em \apj} {\bf 709} (2010) 286--300,
[\href{http://xxx.lanl.gov/abs/0912.0261}{{\tt arXiv:0912.0261}}]

\bibitem{ps}
W.~H. {Press} and P.~{Schechter},
{\it Formation of Galaxies and Clusters of Galaxies by Self-Similar Gravitational Condensation},
{\em \apj} {\bf 187} (1974) 425--438.

\bibitem{st}
R.~K. {Sheth} and G.~{Tormen},
{\it Ellipsoidal collapse and an improved model for the number and spatial distribution of dark matter haloes}.
{\em \mnras} {\bf 308} (1999) 119--126,
[\href{http://xxx.lanl.gov/abs/astro-ph/9901122}{{\tt astro-ph/9901122}}].

\bibitem{jenk}
A.~{Jenkins}, C.~S. {Frenk}, S.~D. M. {White}, J.~M. {Colberg}, S.~{Cole}, 
A.~E. {Evrard},  H.~M. P. {Couchman}, N.~{Yoshida}, 
{\it The mass function of dark matter haloes},
{\em \mnras} {\bf 321} (2001) 372--384,
[\href{http://xxx.lanl.gov/abs/astro-ph/0005260}{{\tt astro-ph/0005260}}].

\bibitem{reed}
D.~{Reed}, J.~{Gardner}, T.~{Quinn}, J.~{Stadel}, M.~{Fardal}, G.~{Lake},
F.~{Governato},
{\it Evolution of the mass function of dark matter haloes}  
{\em \mnras} {\bf 346} (2003) 565--572
[\href{http://xxx.lanl.gov/abs/astro-ph/0301270}{{\tt astro-ph/0301270}}].

\bibitem{warren}
M.~S. {Warren}, K.~{Abazajian}, D.~E. {Holz}, L.~{Teodoro},
{\it Precision Determination of the Mass Function of Dark Matter Halos}, 
{\em \apj} {\bf 646} (2006) 881-885,
[\href{http://xxx.lanl.gov/abs/astro-ph/0506395}{{\tt astro-ph/0506395}}].

\bibitem{courtin}
J.~{Courtin}, Y.~{Rasera},  J.--M.~{Alimi}, P.~S. {Corasaniti}, V.~{ Boucher}, 
A.~{Füzfa},
{\it Imprints of dark energy on cosmic structure formation - II. Non-universality of the halo mass function}, 
{\em \mnras} {\bf 410} (2011) 1911--1931,
[\href{http://xxx.lanl.gov/abs/1001.3425}{{\tt arXiv:1001.3425}}].

\bibitem{bardeen}
J.~M. {Bardeen}, J.~R. {Bond}, N.~{Kaiser}, A.~S. {Szalay}, 
{\it The statistics of peaks of Gaussian random fields},
{\em \apj} {\bf 304} (1986) 15--61.

\bibitem{holder}
G.~{Holder}, Z.~{Haiman}, J.~{Mohr},
{\it Constraints on $\Omega_m$, $\Omega_{\Lambda}$, and $\sigma_8$ from Galaxy Cluster Redshift Distributions},
{\em \apj} {\bf 560} (2001) L111--L114,
[\href{http://xxx.lanl.gov/abs/astro-ph/0105396}{{\tt astro-ph/0105396}}].

\bibitem{planck}
Planck collaboration,
{\it Planck 2013 results. XVI. Cosmological parameters},
submitted to {\em \aap} (2013) ,
[\href{http://xxx.lanl.gov/abs/1303.5076}{{\tt arXiv:1303.5076}}].

\bibitem{CFHT}
C.~{Heymans} et al.,
{\it CFHTLenS tomographic weak lensing cosmological parameter constraints: Mitigating the impact of intrinsic galaxy alignments}.
{\em \mnras} {\bf 432} (2013) 2433--2453,
[\href{http://xxx.lanl.gov/abs/1303.1808}{{\tt arXiv:1303.1808}}].

\bibitem{burenin}
R.~A. {Burenin},  A.~A. {Vikhlinin},
{\it Cosmological parameters constraints from galaxy cluster mass function measurements in combination with other cosmological data},
AstL {\bf 38} (2012) 347--363,
[\href{http://xxx.lanl.gov/abs/1202.2889}{{\tt arXiv:1202.2889}}].

\bibitem{sereno}
M.~{Sereno} and G.~{Covone},
{\it The mass-concentration relation in massive galaxy clusters at redshift $\sim 1$}.
{\em \mnras} {\bf 434} (2013) 878--887,
[\href{http://xxx.lanl.gov/abs/1306.6096}{{\tt arXiv:1306.6096}}].

\bibitem{zentner}
A.~R. {Zentner}, D.~H. {Rudd}, W.~{Hu},
{\it Self Calibration of Tomographic Weak Lensing for the Physics of Baryons to Constrain Dark Energy}.
{\em \prd} {\bf 77} (2008) 043507,
[\href{http://xxx.lanl.gov/abs/0709.4029}{{\tt arXiv:0709.4029}}].

\bibitem{eis}
D.~J. {Eisenstein}, W.~{Hu}, M.~{Tegmark},
{\it Cosmic Complementarity: Joint Parameter Estimation from CMB Experiments and Redshift Surveys}.
{\em \apj} {\bf 518} (1999) 2--23,
[\href{http://xxx.lanl.gov/abs/astro-ph/9807130}{{\tt astro-ph/9807130}}].

\bibitem{hujain}
W.~{Hu} \& B.~{Jain},
{\it Joint Galaxy--Lensing Observables and the Dark Energy}.
{\em \prd} {\bf 70} (2004) 043009,
[\href{http://xxx.lanl.gov/abs/astro-ph/0312395}{{\tt astro-ph/0312395}}].

\bibitem{hukra}
W.~{Hu} \& A.~V. {Kravtsov},
{\it Sample Variance Considerations for Cluster Surveys}.
{\em \apj} {\bf 584} (2003) 207--715,
[\href{http://xxx.lanl.gov/abs/astro-ph/0203169}{{\tt astro-ph:0203169}}].

\bibitem{takada}
M.~{Takada} \& S.~{Bridle},
{\it Probing dark energy with cluster counts and cosmic shear power spectra: including the full covariance}.
NJPh {\bf 9} (2007) 446,
[\href{http://xxx.lanl.gov/abs/0705.01631}{{\tt arXiv:0705.0163}}].

\bibitem{valageas}
P.~{Valageas}, N.~{Clerc}, F.~{Pacaud}, M.~{Pierre},
{\it Covariance matrices for halo number counts and correlation functions}.
{\em \aap} {\bf 536} (2011) A95 ,
[\href{http://xxx.lanl.gov/abs/1104.4015}{{\tt arXiv:1104.4015}}].

\bibitem{lima}
M.~ {Lima} \& W.~{Hu},
{\it Self--calibration of cluster dark energy studies: Counts in cells}.
{\em \prd} {\bf 70} (2004) 043504,
[\href{http://xxx.lanl.gov/abs/astro-ph/0401559}{{\tt astro-ph/0401559 }}].

\bibitem{masamune}
M.~{Oguri},
{\it Self-Calibrated Cluster Counts as a Probe of Primordial Non-Gaussianity}.
{\em \prl} {\bf 102} () 211301,
[\href{http://xxx.lanl.gov/abs/0905.0920}{{\tt arXiv:0905.0920}}].

\bibitem{wechsler}
R.~H. {Wechsler}, A.~R. {Zentner}, J.~S. {Bullock}, A.~V. {Kravtsov}, B.~{Allgood},
{\it The Dependence of Halo Clustering on Halo Formation History, Concentration, and Occupation}.
{\em \apj} {\bf 652} (2006) 71--84,
[\href{http://xxx.lanl.gov/abs/astro-ph/0512416}{{\tt astro-ph/0512416}}].

\bibitem{jingsu}
Y.~P. {Jing}, Y.~{Suto}, H.~J. {Mo},
{\it The dependence of dark halo clustering on the formation epoch and the concentration parameter}.
{\em \apj} {\bf 657} (2007) 664--668,
[\href{http://xxx.lanl.gov/abs/astro-ph/0610099}{{\tt astro-ph/0610099}}].

\bibitem{postman}
M.~{Postman}, L.~M. {Lubin}, J.~E. {Gunn}, J.~B. {Oke}, J.~G. {Hoessel}, D.~P. {Schneider}, J.~A. {Christensen},
{\it The Palomar Distant Clusters Survey. I. The Cluster Catalog}.
{\em \aj} {\bf 111} (1996) 615,
[\href{http://xxx.lanl.gov/abs/astro-ph/9511011}{{\tt astro-ph/9511011}}].

\bibitem{koester}
B.~P. {Koester}, T.~A. {McKay}, J.~{Annis}, R.~H. {Wechsler}, A.~{Evrard}, L.~{Bleem}, M.~{Becker}, D.~{Johnston}, E.~{Sheldon}, R.~{Nichol}, C.~{Miller}, R.~{Scranton}, N.~{Bahcall}, J.~{Barentine}, H.~{Brewington}, J.~{Brinkmann}, M.~{Harvanek}, S.~{Kleinman}, J.~{Krzesinski}, D.~{Long}, A.~{Nitta}, D.~{Schneider}, S.~{Sneddin}, W.~{Voges}, D.~{York}, SDSS collaboration,
{\it A MaxBCG Catalog of 13,823 Galaxy Clusters from the Sloan Digital Sky Survey}.
{\em \apj} {\bf 660} (2007) 239--255,
[\href{http://xxx.lanl.gov/abs/astro-ph/0701265}{{\tt astro-ph/0701265}}].

\bibitem{milke}
M.~{Milkeraitis}, L.~{Van Waerbeke}, C.~{Heymans}, H.~{Hildebrandt}, J.~P. {Dietrich}, T.~{Erben},
{\it 3D-Matched-Filter galaxy cluster finder - I. Selection functions and CFHTLS Deep clusters}.
{\em \mnras} {\bf 406} (2010) 673--688,
[\href{http://xxx.lanl.gov/abs/0912.0739}{{\tt arXiv:0912.0739}}].

\bibitem{bella}
F.~{Bellagamba}, M.~{Maturi}, T.~{Hamana}, M.~{Meneghetti}, S.~{Miyazaki}, L.~{Moscardini},
{\it Optimal filtering of optical and weak lensing data to search for galaxy clusters: application to the COSMOS field}.
{\em \mnras} {\bf 413} (2011) 1145--1157,
[\href{http://xxx.lanl.gov/abs/1006.0610}{{\tt arXiv:1006.0610}}].

\bibitem{campanelli}
L.~{Campanelli}, G.~L. {Fogli}, T.~{Kahniashvili}, A.~{Marrone}, B.~{Ratra},
{\it Galaxy cluster number count data constraints on cosmological parameters},
(2011),
[\href{http://xxx.lanl.gov/abs/1110.2310}{{\tt arXiv:1110.2310}}].

\bibitem{cardonemc}
V.~{Cardone}, S.~{Camera}, M.~{Sereno}, G.~{Covone}, R.~{Maoli}, 
R.~{Scaramella},
(2014), in preparation.

\end{thebibliography}
\end{document}